\documentclass[aps,preprint,tightenlines]{revtex4}
\usepackage{amssymb}

\usepackage{graphicx}

\begin{document}

\bigskip

\bigskip

{\large {\bf ANGULAR DEPENDENT VORTEX DYNAMICS}}

{\large {\bf IN SUPERCONDUCTORS WITH COLUMNAR DEFECTS}}

\bigskip

\bigskip

\bigskip

\hspace{2cm} {\bf LEONARDO CIVALE}

\hspace{2cm} Superconductivity Technology Center,

\hspace{2cm} Los Alamos National Laboratory, Los Alamos, NM 87545, USA.

\bigskip

\hspace{2cm} {\bf ALEJANDRO V. SILHANEK}

\hspace{2cm} Laboratorium voor Vaste-Stoffysica en Magnetisme,

\hspace{2cm} K. U. Leuven, Celestijnenlaan 200D, B-3001 Leuven, Belgium.

\bigskip

\hspace{2cm} {\bf GABRIELA PASQUINI}

\hspace{2cm} Laboratorio de Bajas Temperaturas, Departamento de F\'{\i}sica,  

\hspace{2cm} Facultad de Ciencias Exactas y Naturales,

\hspace{2cm} Universidad de Buenos Aires, Buenos Aires, Argentina.


\bigskip

\bigskip

\bigskip

\bigskip



\section{Introduction}

\subsection{Motivation}

The introduction of aligned columnar defects (CD) in high temperature
superconductors (HTSC) was initially motivated\cite{civale91a} by the technological
objective of increasing the discouragingly low vortex pinning of the early
samples of these compounds. Initial results clearly showed\cite{civale91a,konczy91}
that CD were extremely effective in improving the critical current density ($J_{c}$)
(particularly at high temperatures and magnetic fields) and in enhancing the 
irreversibility field ($H_{irr}$). In addition, it soon became 
clear\cite{nel-vin,blatter94} that good quality single crystals with CD
were excellent ``model systems'' to explore the complexities of vortex
dynamics in HTSC. The combination of both motivations produced a very large
interest on the topic\cite{civalereview}, which remains active after more 
than a decade of research.

Vortex pinning in type II superconductors arises\cite{blatter94,campbell1} from the presence of
crystalline defects, which produce a spatially inhomogeneous superconducting
order parameter $\psi (\mathbf{{r})}$, thus inducing a position dependent
vortex energy. Examples of those defects are dislocations, chemical doping,
vacancies, non superconducting particles, voids, etc. The calculation of
the expected $J_c$ as a function of temperature ($T$) and magnetic induction 
($\bf B$) is a very complex problem\cite{blatter94,campbell1}, as it 
depends nontrivially on a large number of
variables such as the nature of the defects, their size, shape and density,
as well as on the temperature and magnetic field dependent elastic
properties of the vortex array, which in turn are determined by the
superconducting parameters of the material. If the material and/or the
defects are anisotropic, pinning will additionally depend on the orientation
of the applied magnetic field $\bf H$.

A particularly important type of anisotropic pinning is that produced by
extended parallel pinning potentials, commonly known as correlated pinning\cite{nel-vin}.
In HTSC, common examples of correlated pinning structures are the
CD, the twin boundaries\cite{grigorieva92,fleshler93,li93,oussena96,zhukov97b},
and the layered crystallographic structure of these
materials that produces the intrinsic pinning\cite{tachiki89,kwok91}. Although each one has its
distinctive characteristics (CD are linear while the other two are planar,
intrinsic pinning potentials are spatially periodic while the separation
between CD and twins is random), they share several important similarities
in their pinning properties, particularly in regards to the vortex structures 
that occur\cite{nel-vin,blatter94} as a function of the orientation of 
$\bf{H}$. In all cases, pinning is maximum when the vortices are parallel 
to the defects. They all exhibit a lock-in phase: when the angle $\varphi$ between 
$\mathbf{H}$ and the defects is smaller than a certain lock-in angle $\varphi_L$, the
vortex cores remain locked into the pinning potentials. For $\varphi >
\varphi_L$, vortices form staircases, with core segments locked into the
defects connected by weakly pinned kinks.

Different mechanisms can produce aligned linear defects capable of pinning vortices in a superconductor. 
In YBa$_{2}$Cu$_{3}$O$_{7}$ (YBCO), for instance, pinning by dislocations\cite{mannhart,dam} and by the lines occurring at the intersection of two perpendicular families of twins\cite{Boris} has been explored. However, we will restrict the term ``columnar defect" to those produced by irradiation with very energetic heavy-ions. Several features set them apart from the rest. First, they can produce a much larger $J_c$ than any other types of linear defects (as discussed in section I.C. below, this is mainly due to the fact that CD have the appropriate diameter). Second, their density can be easily controlled. Third, they can be introduced at an arbitrary crystalline orientation. This is a unique advantage of the CD as a tool to explore vortex dynamics, as it allows to deconvolute the features arising from their uniaxial pinning from those due to anisotropy and other correlated pinning mechanisms. We also note that by ``aligned" CD we do not mean that all the tracks are perfectly parallel, as will be discussed in the next three subsections. We just use this term as opposite to CD purposely generated with high angular dispersion\cite{kreiselmeyer,lia-splay} (splay) or even randomly oriented\cite{fleischer89,lia-fission}.

The purpose of this review is to discuss the angular dependence of the
vortex dynamics in type II superconductors with aligned CD. The problem is rather
complex, and some aspects still require further investigation. Our main goal is to identify and characterize the various angular regimes, both statically and dynamically. To that end we will use and combine results obtained by dc magnetization and ac susceptibility measurements. We will show that both the crystalline anisotropy and geometrical anisotropy (sample shape) play an important role. Most of our studies have been performed on high quality twinned YBCO single crystals prepared using a flux growth technique \cite{Paco1}. In this compound, the combined effects of the three
types of correlated pinning mechanisms and the anisotropy produce a rich variety of vortex structures and dynamic responses. Complementary studies were performed on high quality single crystals\cite{Bolle93} of the less anisotropic conventional superconductor NbSe$_{2}$, with $T_c = 7.3$ K, coherence length $\xi =77$ {\AA}, and penetration depth $\lambda = 690$ {\AA}, and having the $c$ axis parallel to the thinnest dimension. Magnetization and susceptibility measurements were performed at the Centro At\'{o}mico Bariloche and Centro At\'{o}mico Constituyentes respectively, both dependent on the Atomic Energy Commission of Argentina.

This review is organized as follows: In the remaining of section I, we briefly describe the generation of CD by heavy-ion irradiation (section I.B), and we summarize the basic vortex dynamics for superconductors with aligned CD (the case $\bf H \parallel$ CD is discussed in I.C., and the angular dependence in I.D.). In section II we present an overview of the temperature, field and angular dependence of the irreversible magnetization in samples with
columnar defects and analyze the influence of correlated defects on the
pinning properties of vortex lines. First (in section II.A.) we discuss the origin of
the irreversible magnetization in superconductors and introduce our
measurement procedures. In section II.B. we focus on the angular regimes
associated with different vortex configurations as the external field is
tilted away from the CD orientation. In section II.C. we show that the
sample geometry and the material anisotropy are competing mechanisms that determine
the vortex orientation at low fields. 
In section III we use our ac susceptibility measurements to explore the vortex 
dynamics in YBCO with CD in the vicinity of the solid-liquid boundary. In section 
III.A. we introduce the experimental procedure. In section III.B. we study the 
angular dependence of the ac response with the scope to identify the various angular 
regimes and to investigate which is the main source of pinning at different 
orientations of a low DC field. In section III.C. we explore the dynamic
response of vortices aligned with the CD over a wide range of current densities and
excitation frequencies, and we construct the dynamic phase diagram of the system 
in the temperature vs ac field plane. In section III.D. we extend that study 
to different angles between the applied field and the tracks.


\subsection{The generation of CD by heavy-ion irradiation}

The generation of CD by irradiation with very energetic heavy ions has been
extensively discussed in the literature (for a review see for
example ref. \onlinecite{civalereview} and references therein). Here we will only
present a very brief summary of the basic concepts.

Particle irradiation is a standard way to introduce defects in a solid.
Incident particles transfer energy to a solid\cite{summers89,hensel90,kirk92} 
by direct collisions with lattice atoms (nuclear or nonionizing energy loss, 
$S_n$), and by ionization or electronic excitations (electronic or ionizing 
energy loss, $S_e$). In the case of irradiation with electrons, protons and 
light ions with energies up to a few MeV, almost all the energy is transferred 
via nonionizing energy loss\cite{summers89}. This results in defect formation 
through displacement of the primary recoil atom, and through the cascade of 
collisions that this atom produces if its energy is high enough\cite{kirk92}. 
The size of the damaged region increases with
the primary recoil energy transfer. Electron irradiation (up to a few MeV)
can only produce point defects (vacancy-interstitial pairs or Frenkel pairs),
while protons and light ions generate defects up to several tens of {{\AA}} in
size. Fast neutrons also generate cascade defects, with diameters in the
range of 50 to 100 {\AA}. In all cases, these defects are randomly
distributed.

As heavier and more energetic ions are considered, $S_{e}$ grows and the
electronic energy loss becomes dominant\cite{hensel90}. For instance, for our typical
irradiations using $\sim$ 300 MeV Au, $S_{e}\sim 100S_{n}$, while for ion energies
in the GeV range the relation can be as large\cite{fuchs87} as $S_{e}\sim 2000S_{n}$. In
this limit each incident ion forms a cylinder of amorphyzed material along
its path, around 40 to 100 {{\AA}} in diameter, known as a latent track. The
heavier and more energetic the ion, the more continuous, homogeneous and
parallel are the tracks. In fact, as $S_{e}$ increases the defects evolve\cite{fuchs87,studer92} 
from isolated uncorrelated spheres to aligned spheres, then to elongated
aligned but disconnected defects, which then coalesce into discontinuous
inhomogeneous tracks, which eventually become continuous and finally
homogeneous. However, as will be pointed out below, perfectly aligned
and homogeneous defects are not necessarily the best suited for high vortex
pinning.

The $S_{e}$ values separating the various regimes depend on the material.
The thresholds are lowest in insulators, while in the other extreme it is
impossible to create continuous tracks in good metals. For the case of YBCO,
irradiation with 173 MeV Xe ($S_{e}\sim 1.2$ KeV/{\AA}), for instance, does not
generate defects capable to produce uniaxial pinning. In this material, the
threshold for the formation of continuous inhomogeneous tracks is\cite{marwick} $S_{e}\sim
2$ KeV/{\AA}, somewhat below the $2.7$ KeV/{\AA} that occurs at the entry surface
for the 580 MeV $Sn^{30+}$ used\cite{civale91a} in the initial studies. Heavier ions such as
Au and Pb in the GeV energy range, which have $S_{e}$ as high\cite{marwick} as 3.5 to 4.5 
KeV/{\AA}, result in homogeneous tracks. All these $S_{e}$ values are
extremely large, so incident ions lose energy very fast as they penetrate
deeper into the target. As they do, at some point $S_{e}$ starts to decrease, so the morphology of
every track changes as a function of depth, until the ion is eventually
stopped. Typical range of continuous tracks formation is about $10~\mu$m for
our 300 MeV Au, and can be as large as $\sim 100~\mu$m for energies in the
several GeV range. As ions penetrate deeper into the material, the angular
dispersion of the tracks also increases\cite{civale94}. This is mainly due to Rutherford
scattering of the incident ions, associated to the small but nonzero nuclear
energy loss. A completely different approach based on fission tracks can be
used\cite{fleischer89,lia-fission} to generate randomly oriented CD.

Irradiations reported in this review were performed at room
temperature at the TANDAR accelerator facility. A group of samples was
irradiated with 291 MeV Au$^{27+}$ ions and another one
with 220 MeV Sn$^{22+}$ ions. In both cases, the energy deposition
rate is greater than the threshold for continuous latent track formation in the first 
$9-10~\mu$m. So we only used crystals of thickness $\le 10~\mu$m.

\subsection{Vortex dynamics in superconductors with CD: basic results for $\mathbf{{H}\parallel}$ CD}

In this subsection we present a brief summary of the basic results of vortex dynamics when the applied magnetic field is parallel to the CD. More detailed theoretical analyses can be found in refs. 3 and 4.

The basic reason why CD are very effective pinning centers when the vortex
direction is parallel to them is quite simple: each CD can in principle
confine the whole length of the vortex core, thus producing the maximum
possible pinning energy without any cost in elastic energy due to vortex
bending. This contrasts with the case of randomly distributed defects, where
only a fraction of the core length is pinned, and additionally vortices must
zig-zag among defects in order to get pinned, thus paying an additional
price in terms of vortex elastic energy.

Let's start with a rough estimate. At low temperatures and for an isolated
vortex (\textit{i.e.}, at low enough fields) the pinning energy per unit length of a 
CD (assuming $\psi=0$ inside it) is $\varepsilon_p \sim (H_c^2/8\pi)\pi b^2$, 
where $H_c$ is the thermodynamic critical field and $b$ is the smallest of the 
defect radius $r_{D}$ and the core radius $\sqrt{2}\xi$ (here $\xi$ is the 
superconducting coherence length).
The pinning force per unit length will be of the order of $F_p \sim \varepsilon_p/\xi$, 
and the critical current density $J_c = cF_p/\Phi_0$, where $\Phi_0$ is
the flux quantum and $c$ is the speed of light. Clearly, the most convenient
case is $r_{D} \sim \sqrt{2}\xi$, so the highest $F_p$ is achieved without
removing more superconducting material than necessary. In this case we
obtain $J_c \sim J_0/2$, where $J_0 = cH_c/3\sqrt{6}\pi\lambda$ is the
depairing current and $\lambda$ is the superconducting penetration depth for ${\bf H} \parallel$ $c$ axis. A
more exact calculation gives\cite{nel-vin,blatter94} $J_c \sim \left(3\sqrt{3}/4\sqrt{2} \right) J_0
\sim J_0$, showing that, at least in principle, CD can produce a $J_c$
pretty close to the maximum theoretically achievable value. It is a fortunate 
fact that the CD produced by heavy-ion irradiation have a radius 
$\sim 20 - 50$ {\AA} that matches very well with the vortex core radius 
$\sqrt{2}\xi \sim 25$ {\AA} of HTSC. On the other hand, the reality is that the largest $J_c$'s 
measured\cite{civalereview}, at low temperatures in YBCO crystals with CD are $\sim 4-5 \times 10^7$ A/cm$^2$. Although these values are very high, they are far from the ideal limit $\sim J_0(T=0) \sim 3 \times 10^8$ A/cm$^2$. This may be due to a variety of factors related to the morphology of the tracks, such as its inhomogeneities, the degree of depression of $\psi$, the sharpness of the normal/superconducting boundary, etc. It is thus useful to introduce\cite{krusin96a} a dimensionless \textit{efficiency factor} $\eta \leq 1$ in the pinning estimates, which accounts for all the uncertainties mentioned.

Several additional factors reduce $\varepsilon_p$ (and thus $J_c$) as temperature
increases\cite{nel-vin,blatter94}. First, $H_c(T)$ decreases with T. Second, as $\xi(T)$ increases
with T a crossover from $r_D>\sqrt{2}\xi$ to $r_D<\sqrt{2}\xi$ may occurs. The 
third and most important factor is due to thermal fluctuations and is
known as entropic smearing. The transverse localization length of the vortex cores, which is determined by the root mean square amplitude of the
thermal fluctuations, grows as $k_BT$, thus reducing the vortex free energy $F=U-TS$ due
to the increase of the entropy $S$ arising from the delocalization. As a
consequence, the effective potential well associated to the CD becomes
``shallow" and the pinning energy decreases by a factor $f(x)$,
where $f(x) \sim \exp(-x)$ for $x>1$, according to the long range
nature of the CD-vortex interaction\cite{blatter94}. Here $x=T/T_{dp}$ and 
$T_{dp}$ is called the depinning temperature. In the very useful
analogy to 2D-bosons\cite{nel-vin}, this corresponds to the decrease in binding energy of
a quantum well due to the zero-point fluctuations. The entropic smearing
effect has been experimentally observed in YBCO, and it was found\cite{krusin96a,civale96a} that $T_{dp}\sim40$ K.
The combination of all the factors mentioned above produce an $\it {effective}$ pinning energy per unit length

\begin{equation}
\varepsilon _p(T)=\eta \frac{\varepsilon _{0}}{2}\ln \left( 1+\frac{r^{2}}
{2\xi ^{2}}\right) \times f(x)  \label{eq:pinenergy}
\end{equation}
where $\varepsilon_{0}(T)=\left( \Phi _{0}/4\pi \lambda \right) ^{2}$ is the vortex energy scale, and the logarithmic factor provides a convenient interpolation\cite{blatter94} between the $r_D>\sqrt{2}\xi$ and $r_D<\sqrt{2}\xi$ cases.

As the vortex density increases, more CD will be occupied. The natural
characteristic field in the problem is\cite{civale91a} the matching field $B_{\Phi }=n\Phi
_{0}$, where $n$ is the areal density of CD. For $B\gg B_{\Phi }$ there are
many more vortices than CD, so all the CD are occupied and only a small
fraction $\sim B/B_{\Phi }$ of the flux lines are actually trapped, the rest
of them being pinned only as a result of vortex-vortex interactions. We thus
expect $J_{c}$ to decrease fast with $B$ in this regime, as indeed observed\cite{civale91a,civalereview,krusin96a}.
When $B\ll B_{\Phi }$ each flux line will be confined into a CD.
However, as $B$ increases it may occur that, although there is a CD
available, it will be energetically unfavorable for the vortex to occupy it
because it is too close to another vortex, so the gain in pinning energy
will be overcompensated by the increase in the elastic energy associated to
the vortex-vortex repulsion. The characteristic field where this effect sets
in is called the {\it accommodation field}, 
$B_{a}(T)\propto \left(\varepsilon_{p}(T)/\varepsilon _{0}\right) B_{\Phi }$, 
which represents the crossover between single vortex pinning where vortex
interactions are negligible and collective pinning where they become
significant\cite{nel-vin,blatter94,krusin96a}. At low temperatures $B_{a}$ approaches $B_{\Phi }$, but above 
$T_{dp}$ it decreases due to the reduction in $\varepsilon_{p}(T)$. Above a temperature $T_{dl}>T_{dp}$ the
localization length becomes larger than the average distance between tracks, so
vortices become collectively pinned by several tracks and $B_{a}(T)$ decreases faster.
We have estimated\cite{Gabi2} that in YBCO $T_{dl}\sim 70$K.


Thermal fluctuations are also responsible for the large flux creep in HTSC.
In the presence of CD and below $B_{a}(T)$, initial relaxation proceeds via
nucleation and expansion of half loops\cite{nel-vin,blatter94,niebied}. This is a glassy regime, in the
sense that the current density dependent {\it activation energy} $U(J)$ increases as $J^{-1}$ with decreasing
current density. Below a certain $J$, a crossover to a creep regime
dominated by double kinks occurs\cite{niebied}. These excitations involve the transfer of
a flux line from one CD to another one, and once a double kink is formed its
further expansion is energetically favorable, thus relaxation in this regime
is fast and {\it nonglassy}. Below a still lower $J$, the expansion of double
kinks is precluded by the dispersion in the pinning energy of the CD.
Relaxation now evolves through another type of excitations called
superkinks. This regime, analogous to the variable range hopping
conductivity in doped semiconductors, is again glassy, with $U(J)$
diverging as $J^{-1/3}$. For $B>B_{a}(T)$ intervortex
interactions are relevant at all current densities, and the relaxation takes
place via collective excitations involving vortex bundles. A description of
the relaxation regimes in that case can be found in Ref. \onlinecite{blatter94}.

An interesting implication of the above analysis is that identical and
perfectly parallel CD are not the ideal configuration to sustain large
persistent current densities, because they allow for the fast, nonglassy
relaxation of double kinks\cite{nel-vin,blatter94,niebied}. In fact, it has 
been theoretically predicted\cite{hwa}, and experimentally 
demonstrated\cite{lia-splay,civale94}, that larger persistent currents can be
obtained in the case of CD having a small angular dispersion or splay. The
case of very large angular splay is also interesting and has been explored
by generating randomly oriented fission tracks\cite{fleischer89,lia-fission}.

A related phenomenon is the matching effect. When $B=B_{\Phi }$, the
available phase space for transfer of vortex cores between two CD should be
strongly limited. Thus, creep rate should be reduced and the field
dependence of the persistent current should exhibit a peak at $B_{\Phi}$.
However, this matching effect is generally not observed\cite{civale91a,civalereview}. 
The reason is that usually the energy and angle dispersion of the CD reduces 
the creep at all fields\cite{niebied-matching}, thus masking this phenomenon. Matching effects have only been
observed\cite{niebied-matching,mazilu-matching} in very thin samples containing CD with very low energy and angle
dispersion.

\subsection{Angular dependence: Bose glass picture}

Our studies of the angular dependence of vortex pinning in HTSC with CD 
have revealed a richer variety of phenomena and pinning regimes than originally 
expected. As we will extensively discuss in the following sections, this 
is due to a number of additional ingredients such as influence of other types of correlated disorder, energy dispersion of the CD and misalignments between $\bf{B}$ and $\bf{H}$ due to both mass anisotropy and sample shape. But first, in this subsection we briefly summarize the expectations within the basic Bose Glass scenario\cite{nel-vin,blatter94} for a single family of identical and perfectly parallel CD. In principle, this picture also applies (with minor differences) to twins and intrinsic pinning.

First, when the angle between the applied field $\mathbf{H}$ and the correlated defects is smaller than a certain 
$\varphi_{L}$, it is energetically favorable for the vortices to remain locked into the defects. This is related to the concept of {\it transverse Meissner effect}. The lock-in angle is

\begin{equation}
\varphi _{L}\simeq \frac{4\pi \sqrt{2\varepsilon _{l}\varepsilon _{p}}}{\Phi _{0}H}  \label{eq:lockin}
\end{equation}
where $\varepsilon _{l}$ is the vortex line tension. In isotropic superconductors $\varepsilon _{l} = \varepsilon _{0}\ln \kappa$, but in the anisotropic case the situation is rather complex. In our experiments the appropriate line tension is that corresponding to in-plane deformations (see pages 1163-1164 in Ref. \onlinecite{blatter94}), 
$\varepsilon _{l}=\left( \varepsilon ^{2}\varepsilon _{0}/\varepsilon (\Theta )\right) \ln \kappa $, where the mass anisotropy $\varepsilon =m_{ab}/m_{c}\ll 1$ and $\varepsilon ^{2}(\Theta )=\cos ^{2}(\Theta )+\varepsilon ^{2}\sin
^{2}(\Theta )$. Note that both $\varepsilon_{l}$ and $\varepsilon_{p}$ decrease with $T$.

For tilt angles larger than $\varphi _{L}$ and smaller than a trapping
angle, $\varphi _{T}=\sqrt{2\varepsilon _{p}/\varepsilon _{l}}$, vortices 
form staircases with segments pinned into different defects and
connected by unpinned or weakly pinned kinks. Beyond $\varphi _{T}$,
vortices become straight lines aligned with the applied field
and unaffected by the correlated nature of the pinning potential. 

Experimentally, the determination of bulk vortex structures for applied fields tilted
with respect to the correlated pinning potential is very difficult.
Most of the existing imaging methods that permit direct observation of the vortices can only detect the flux lines at the
sample surface\cite{yao}. Recent studies have shown that it is possible to resolve individual vortices inside a superconductor with columnar defects by means of Lorentz microscopy\cite{harada} or interference microscopy\cite{tonomura}. Unfortunately, however, the low penetration power of the electron beam imposes a strong restriction in the maximum
thickness of the sample of about $0.5~\mu$m. Thus, in general, in order to
obtain information from the 3D nature of the vortex structure we still have
to rely on indirect techniques like dc-magnetization, ac-susceptibility or
transport measurements\cite{civale91a,kwok92,grigorieva92,fleshler93,li93,klein93,holzapfel93,hardy96,oussena96,zhukov97b,herbsommer98}. 
In this context, it is useful to introduce CD tilted off the crystallographic $c$ axis, as this helps to discriminate the pinning produced by them from that due to twin boundaries and from anisotropy effects.


\section{DC Magnetization: Angular dependent vortex\\ configurations in the solid phase}


\subsection{Introduction to dc-Magnetization Measurements}

The widely used dc magnetization measurements are a powerful
tool to explore the irreversible response of the vortex system deep into the vortex-solid
phases, where electro-transport measurements and ac-susceptibility hardly
access. 
If a superconductor is cooled down in a zero applied field (ZFC experiment) and then an external
field $H>H_{c1}$ is applied, flux-bearing vortices enter through the sample's border until their motion is arrested
by pinning centers. As a consequence, the system achieves an inhomogeneous
flux distribution with a higher density of vortices near the border that
progressively decreases toward the center of the sample. The spatial
variation of the locally averaged field $\mathbf{B}(\mathbf{r})$ gives rise
to supercurrents $\mathbf{J}$ in the sample that, in the stationary state,
accommodate to be exactly the critical current $J_{c}$ everywhere. This is
the so-called \textit{Critical State} regime\cite{campbell1,bean}.

The simplest version of the critical state is the isotropic Bean model\cite{bean}, where $J_c$ is assumed to be field independent and isotropic, so the current density throughout the sample (in the fully penetrated state) is necessarily uniform, and the associated irreversible magnetization $\bf{M_i}$ is proportional to $J_c$ via a geometrical factor. For instance, for a disk of radius $R$ with $\bf{H}$ parallel to its normal $\mathbf{\hat n}$ we have ${\bf M_i}=\left(J_cR/3c\right)\mathbf{\hat n}$. However, even in this simple case the calculation of the field profile is rather complex, and has been the subject of extensive modelling and numerical analysis\cite{daeumling,clem-san,brandtA}. The situation is further
complicated when realistic cases of field dependent and anisotropic critical currents are considered\cite{conner,gyorgy}. 

\begin{figure}[htb]
\centering
\includegraphics[angle=0,width=40mm]{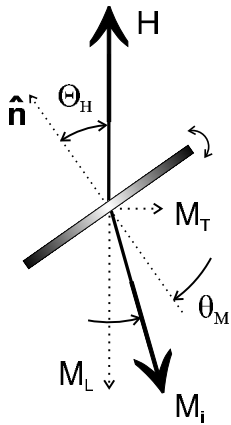}
\caption{{\protect\small Sketch showing the orientation of the irreversible
magnetization $\mathbf{M_i}$ and applied field $\mathbf{H}$ with respect to
the sample normal $\mathbf{\hat n}$.}}
\label{layout}
\end{figure}

If ${\bf H}$ is tilted with respect to $\mathbf{\hat n}$, then $\bf{M_i}$ has two components, $M_\parallel $ and $M_\bot$, parallel and perpendicular to $\mathbf{\hat n}$ respectively (see Fig. \ref{layout}), that in general depend on the various components of $J_c$ and on the sample geometry. In very thin samples (which is our case), since the non-equilibrium screening currents are constrained to flow parallel to the sample surface, there is a large angular range of applied
fields\cite{hellman,zhukov97a} in which $\bf{M_i}$ points almost parallel to $\mathbf{\hat n}$. For an infinite strip of an isotropic superconductor, Zhukov {\it et al.}\cite{zhukov97a} showed that $\bf{M_i}$ remains almost locked to $\mathbf{\hat n}$ as long as the angle $\Theta_H$ between ${\bf H}$ and $\mathbf{\hat n}$ remains smaller than $\Theta_c = \arctan \nu^{-1}$. Here $\nu=\delta/w$, where $\delta$ and $w$ are the thickness and width of the strip respectively, with $\delta \ll w$. For this particular geometry, 

\begin{equation}
M_\bot = \frac{\nu}{6c} J_c \tan \Theta_H~~~M_\parallel = \frac{J_c w}{4c}%
\left( 1- \frac {\nu^2}{3} \tan^2 \Theta_H \right)~~~if~~\Theta_H \leq
\Theta_c
\end{equation}
and the angle $\Theta_M$ between $\mathbf{M_i}$ and $\mathbf{\hat n}$ is
given by,

\begin{equation}
\tan \Theta_M = \frac{M_\bot}{M_\parallel} = \frac{2\nu^2\tan \Theta_H}{%
3-\nu^2\tan^2\Theta_H}\approx \frac{2}{3} \nu^2 \tan \Theta_H
\end{equation}

Typically we deal with samples of dimensions $\delta \sim 15 ~\mu$m and $w \sim
400 ~\mu$m, so $\nu \sim 0.04$ and $\Theta_c = 87^\circ$, \textit{i.e.} $M_\parallel
\gg M_\bot$ and the geometrical factor changes only $0.1 \%$ for $\Theta_H
\leq \Theta_c$. On the other hand, for $\Theta_H \geq \Theta_c$ the
irreversible moment vector suddenly rotates towards the sample plane.

It must be noted that the total field in the sample results from the applied field plus the \textit{self field}
$H_{sf}$ generated by the screening currents. At the center of a thin disk $H_{sf} \sim J_c t$ (rather than $\sim J_c R$) and the shielding currents also create radial fields of $\sim H_{sf}/2$ on the disk surface\cite{daeumling,clem-san,brandtA}. For $H < H_{sf}$ self field effects dominate, thus $J_c(H)$ cannot be easily extracted from $M_i(H)$. We constrain our analysis to the regime $H \gg H_{sf}$ in order to obtain reliable estimations using the Bean model. For instance, at $T = 60$ K for YBCO crystals of $\delta \sim 15 \mu $m with CD, $J_c \sim 10^5$ A/cm$^2$ and hence $H_{sf} \sim 150 $ G. 

In spite of the shortcomings of the Bean model and the necessity of extend
this description to account for anisotropies and
field dependencies of $J_{c}$, several experiments have shown that the
behavior of the irreversible magnetization follows accurately this model\cite
{schusterA,tamegai} justifying its use as a reliable way to determine the
critical current of the system\cite{shantsev}. 

\vspace{0.5 cm}

The magnetization measurements presented in this section were conducted
on a SQUID-based magnetometer Quantum Design MPMS-5S equipped with two sets
of detectors, which allows us to record both the longitudinal $M_L(H)$ and
the transverse $M_T(H)$ components (parallel and perpendicular to $\mathbf{H}$,
respectively) of the total magnetization vector $\mathbf{M}$. The crystals can be rotated \textit{in situ} around an axis
perpendicular to $\mathbf{H}$ (see Fig. \ref{layout}), and they are
carefully aligned with the rotating axis normal to the irradiation plane, in
such a way that the condition $\mathbf{H}\parallel$ tracks could be achieved
within $\sim 1^{\circ }$.


%


Our usual measurement procedure\cite{evidence,anomalous} consists in a ZFC of the sample from above $T_c$ down to
the work temperature $T$ at a given $\Theta_H$. Then, both components $M_L(H)$ and $M_T(H)$ of a magnetization loop are recorded (at fixed $T$ and $\Theta_H$). After that, the sample is rotated to a new desired angle, warmed up above $T_c$ and cooled down in zero field to start a new run. The total magnetization vector so obtained is $\mathbf{M} = \mathbf{M_i} + \mathbf{M}_{eq}$, where $\mathbf{M}_{eq}$ is the equilibrium (reversible) magnetization in the mixed state of the superconductor, plus any contribution from the sample holder. We use the widths of the hyterisis $\Delta M_L(H)$ and $\Delta M_T(H)$ to calculate the modulus $M_i = \frac 1 2 \sqrt{\Delta M^2_L(H)+\Delta M^2_T(H)}$ and the angle $\Theta_M = Atan(\Delta M_T(H)/\Delta M_L(H))$ of $\mathbf{M_i(H)}$. Note that in this way $M_{eq}$ automatically cancels out of the calculation, so it is not necessary to determine it.

We have also developed an alternative method\cite{avila01} which allows us to obtain directly $M_{i}(\Theta _{H})$ by rotating the sample at fixed $H$ and $T$. This procedure is much faster, and its principal advantage is that a finer
grid can be easily obtained in the angular range of interest, thus permitting the exploration of various regimes with significantly improved angular resolution. The complication is that the $M_{eq}$ is not automatically cancelled, thus it must be subtracted (usually, however, in HTSC the strong pinning determines that $M_i \gg M_{eq}$). 
We have solved that complication by appropriate data analysis\cite{avila01}, and we have been able to obtain very good agreement between data from sample rotations and from magnetization loops. A rotation at fixed $H$ is to some extent analogous to a hysteresis loop\cite{prozorov96}. Rotating the sample normal, $\bf \hat n$, towards $\bf H$ increases $H_{\perp}$, which is roughly equivalent to increasing $H$ at $\Theta_H=0^{\circ}$, moving along the lower (diamagnetic) branch of the loop. Decreasing $H_{\perp}$ (either by rotating $\bf n$ away from $\bf H$ or by crossing the $\bf H \parallel \bf c$ condition), is equivalent to reversing the field sweep, thus producing a switch to the other branch of the loop. This is a useful analogy for the analysis of the rotations, although it should not be pushed too far.

Finally, it is important to mention that, due to the large influence of thermal fluctuations on the vortex dynamics in these HTSC compounds, the persistent current $J$ determined through dc magnetization measurements in the typical time scale of SQUID magnetometers ($\sim 30$ sec.) is much smaller than the ``true" critical current density $J_c$.


\subsection{\protect\bigskip Angular regimes}

\subsubsection{Lock-in Phase}

\begin{figure}[htb]
\centering
\includegraphics[angle=0,width=105mm]{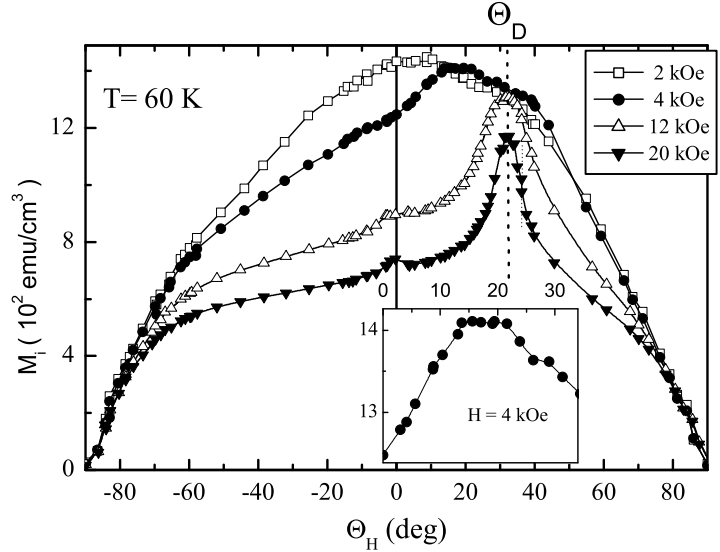}
\caption{{\protect\small Irreversible magnetization $M_i(H)$ as a
function of $\Theta_H$ for several fields, at $T=$ 60 K. Inset: zoom in of
the data of the main panel for $H=4$ kOe around the plateau region.}}
\label{MivsH}
\end{figure}

Figure \ref{MivsH} shows typical $M_{i}$ versus $\Theta_H$ curves at $T=60$ K
for a YBCO single crystal with CD at an angle $\Theta _D = 32^\circ$ away from the
$c$ axis\cite{evidence}. As discussed above, the geometrical factor between $M_{i}$ and $J$ is almost constant for $\Theta_H \le \Theta_c \sim 87^\circ$ in our crystal, thus the vertical axis in Fig. \ref{MivsH} is directly proportional to the persistent $J$ over almost all the angular range. The uniaxial nature of the pinning potential is clearly manifested
as an asymmetric angular response $M_i(\Theta_H) \neq M_i(-\Theta_H)$. At
high fields ($H\geq 10$ kOe) we observe a large peak in the direction of the
tracks $\Theta_{max} \simeq \Theta _D = 32^{\circ }$. At lower fields the
peak becomes broader and transforms into a plateau (the angular range where $%
M_i(\Theta_H) \sim const.$) as well as it \textit{progressively shifts away
from the tracks in the direction of the $c$ axis} ($\Theta_{max} < \Theta _D$%
). Hereafter in this section we will focus in the study of the plateau and
defer the discussion of the shift to the next section.

%

\begin{figure}[htb]
\centering
\includegraphics[angle=0,width=90mm]{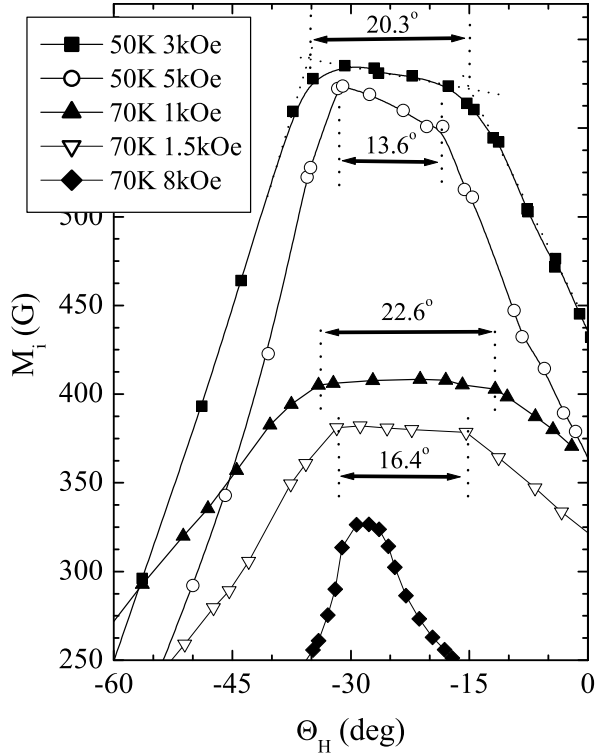}
\caption{$M_i$ versus $\Theta_{max}$ in the region of the plateau for
several fields and temperatures in a ErBCO single crystal.}
\label{plateau}
\end{figure}

The inset of Figure \ref{MivsH} shows a zoom in of the data of the main
panel for $H=4$ kOe, where the plateau is clearly observed. The
constancy of $M_{i}\left( \Theta_H\right)$ indicates that the pinning energy
remains constant and equal to the value at the alignment condition $%
\Theta_B=\Theta _D$. This behavior is a fingerprint of the lock-in phase\cite
{nel-vin}. Moreover, the decrease of $M_{i}$ at the edges of the plateau is
quite sharp, a result consistent with the appearance of kinks, which not
only reduce $J_c$ but also produce a faster relaxation. A few examples of
the observed plateaus for a ErBa$_2$Cu$_3$O$_7$ (ErBCO) single crystal are shown in Figure \ref{plateau}
for several fields and temperatures (some curves have been translated
vertically for clarity).

%

\begin{figure}[htb]
\centering
\includegraphics[angle=0,width=105mm]{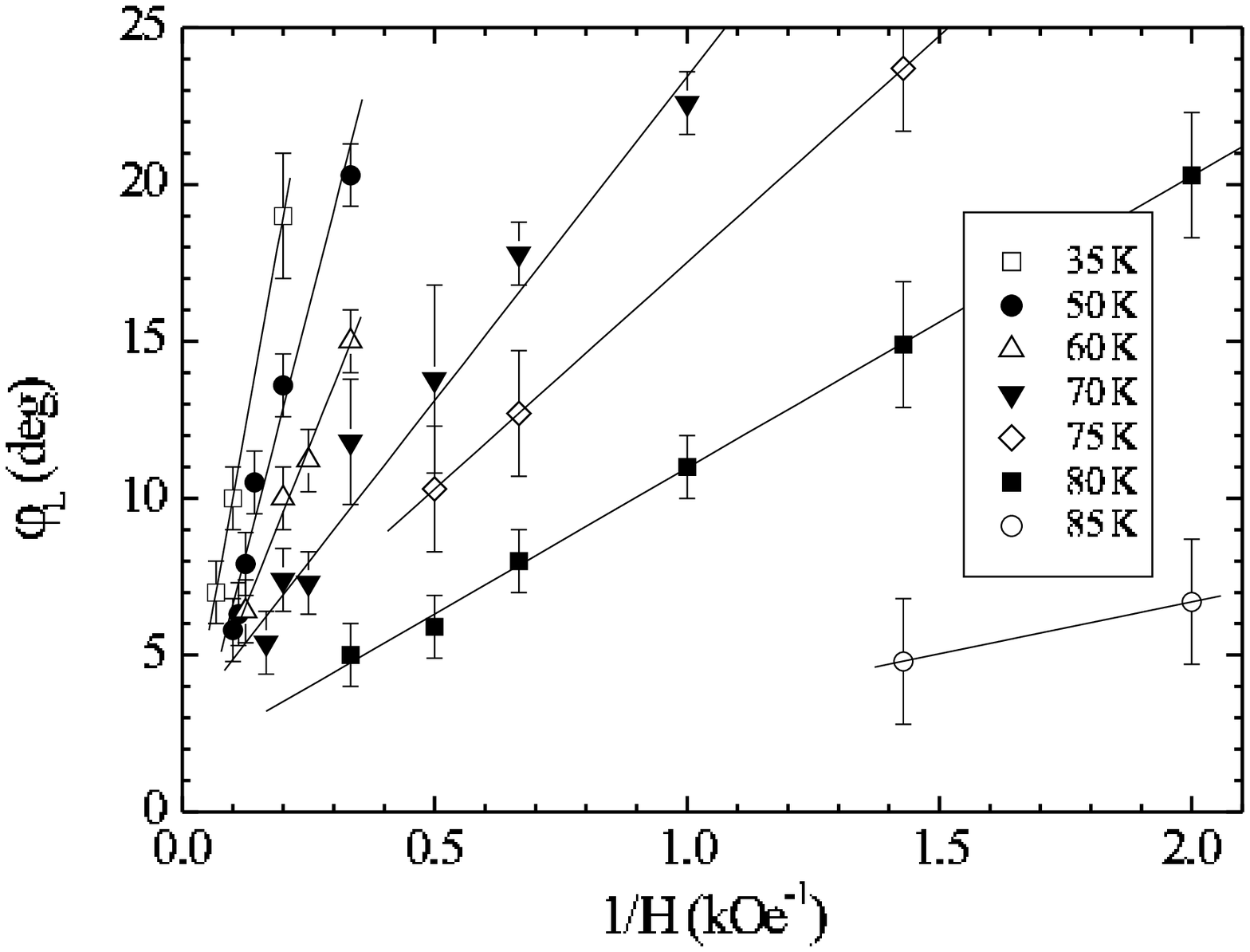}
\caption{Plateau widths $\protect\varphi_L$ vs $H^{-1}$ for several
temperatures. The straight lines are fits according to Eq. (\ref{eq:lockin}).}
\label{lockinvsH}
\end{figure}

The plateau represents the angular range of applied field over which it is
energetically convenient for the vortices to remain locked into the columnar
defects, therefore its angular width is twice the lock-in angle $\varphi_L$.
The results for all measurable plateau widths are plotted as a
function of $H^{-1}$ in Figure \ref{lockinvsH}. This figure clearly
demonstrates the linear dependence of $\varphi_L$ on $H^{-1}$ over the whole range of
temperature and field of our study (as evidenced by the
solid lines which are the best linear fits to the data), in agreement with the Bose-glass prediction [Eq. (\ref{eq:lockin})].

%

\begin{figure}[htb]
\centering
\includegraphics[angle=0,width=105mm]{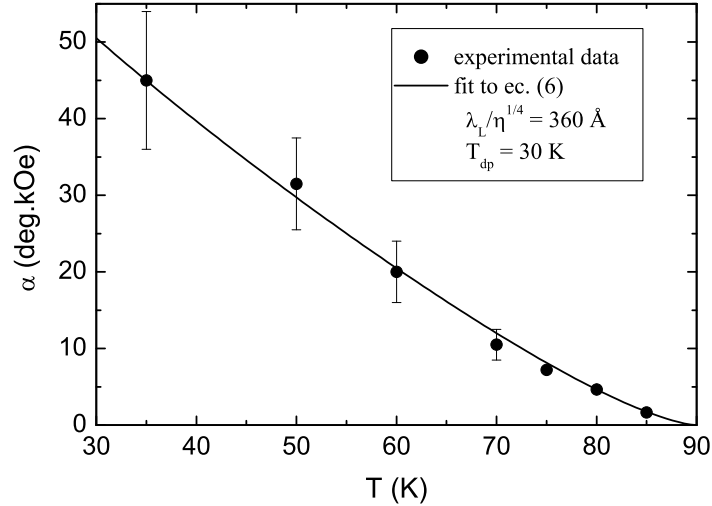}
\caption{Slopes $\protect\alpha(T)=d\protect\varphi_L/d(H^{-1})$ of the
linear fits of Fig.~\ref{lockinvsH}}
\label{lockinvsT}
\end{figure}

To analyze whether Eq. (\ref{eq:lockin}) provides a satisfactory description of the temperature dependence of the lock-in effect, the slopes $\alpha (T)=d\varphi _{L}/d(H^{-1})$ of the linear fits showed in Fig.~\ref{lockinvsH} are plotted 
in Fig.~\ref{lockinvsT} (solid symbols). As expected, $\alpha (T)$ decreases with increasing $T$, reflecting the fact that the lock-in angle at fixed $H$ decreases with $T$ due to the reduction of both, the line tension and the pinning energy. 
From eqs. (\ref{eq:pinenergy}) and (\ref{eq:lockin}) evaluated at the track's direction $\Theta_B=\Theta _{D}$ we obtain

\begin{equation}
\alpha(T) \approx \frac{\Phi_0\varepsilon}{8\pi\lambda^2} \ln\left(1+\frac{%
r_{D}^{2}}{2\xi^2}\right)\times \left[\eta \frac{2\ln\kappa}{\varepsilon(\Theta_D)}
f(x) \right]^{1/2},  \label{eq:slope2}
\end{equation}
where $f(x) \sim \exp(-x)$ for $x>1$.

We can now fit the experimentally determined $\alpha(T)$ using Eq. (\ref
{eq:slope2}). To that end we fix the reasonably well known superconducting
parameters of the material $\varepsilon \approx 1/5$; $\ln\kappa \approx 4$
and $\xi=15${\AA} $/\sqrt{1-t}$ (where $t=T/T_c$). We also assume the usual
two-fluid temperature dependence $\lambda(T)=\lambda_L/2\sqrt{1-t^4}$, where 
$\lambda_L$ is the zero-temperature London penetration depth. The free
parameters are then $T_{dp}$ and the combination $\lambda_L/\eta^{1/4}$. The
best fit, shown in Fig.~\ref{lockinvsT} as a solid line, yields $%
\lambda_L/\eta^{1/4}=360$ {\AA} and $T_{dp}=30$ K.

%

The obtained $T_{dp}$ is smaller, but still reasonably similar to the value $%
\sim 40$ K reported for several YBCO crystals using a completely different
experimental method\cite{krusin96a,civale96a}. This low $T_{dp}$ (well below
the initial theoretical expectations) indicates that the efficiency factor $%
\eta$ is rather small, what is also consistent with the less than optimum $%
J_c$ observed here and in several previous studies. For low matching fields
as that used in the present work, it was estimated\cite{krusin96a} that $%
\eta \sim 0.2 - 0.25$.

The exact value of $\eta$ has little influence in our estimate of $\lambda_L$%
, as it only appears as $\eta^{1/4}$. For $\eta=0.2$ and $\eta=1$ we get $%
\lambda_L=250$ {\AA} and $360$ {\AA} respectively, a factor of 4 to 5 smaller
than the accepted value $\lambda_L \sim 1400$ {\AA}. Zhukov {\it et al.}\cite
{zhukov97b} had reported a similar discrepancy when studying the lock-in
effect by both CD and twin boundaries in YBCO. Similarly, as we will show in
the next section, we also find that the misalignment between $\mathbf{B}$
and $\mathbf{H}$ at low fields can be described satisfactorily using a $%
\lambda_L$ smaller than the accepted value\cite{evidence}. Thus, this
numerical discrepancy appears to be a common result associated to the study
of angular dependencies in YBCO-type superconductors with correlated
disorder that deserves further analysis.

Finally, it is relevant to note that Eq. (\ref{eq:lockin}) was derived for
the \textit{single vortex pinning} regime, which occurs below the
temperature dependent accommodation field $B_a(T)<B_{\Phi }$ (section
I.C.), while a large fraction of the data shown in Fig.~\ref{lockinvsH} lies
above this line, in the \textit{collective pinning} regime. Unfortunately,
to our knowledge there is no available expression for $\varphi _{L}\left(
H,T\right) $ in the collective regime. Blatter {\it et al.}\cite{blatter94} only
argued that collective effects should result in a reduction of the lock-in
angle. The experimental fact is that Eq. (\ref{eq:lockin}) satisfactorily
describes both the temperature and field dependence of $\varphi _{L}$. This
suggests that, at least to a first approximation, collective effects in the
range of our measurements simply result in a different prefactor in Eq. (\ref
{eq:lockin}).


\subsubsection{Trapping Angle}


As was mentioned in Section I.D., it is expected that for $\left| \Theta _{B}-\Theta _{D}\right| > \varphi _{L}$ vortices form staircases. The question that arises here is whether we are able to observe
any clue of the trapping angle $\varphi _{T}$ from the angular dependence of
the irreversible magnetization. For clarity lets concentrate the analysis on
high enough fields such that no plateaus or peak shifts appear.

First, we have to notice that for $\Theta_H>\Theta_{\max }$, the asymmetry 
$M_{i}\left( +\Theta_H\right) >M_{i}\left( -\Theta_H\right) $ in Fig. \ref{MivsH} indicates that
pinning is stronger when $H$ is closer to the tracks than in the
crystallographically equivalent configuration in the opposite side. This
asymmetry demonstrates that at the angle $+\Theta_H$ vortices form
staircases, with segments trapped into the tracks. Second, for $%
\Theta_H<\Theta_{\max }$ we again observe asymmetry, $M_{i}\left(
\Theta_H\right) $ crosses $\Theta_H=0$ with positive slope, indicating that
pinning decreases as $H$ is tilted away from the tracks. Thus, we can
conclude that staircases extend at least beyond the $c$ axis into the $%
\Theta_H<0$ region.

\begin{figure}[htb]
\centering
\includegraphics[angle=0,width=100mm]{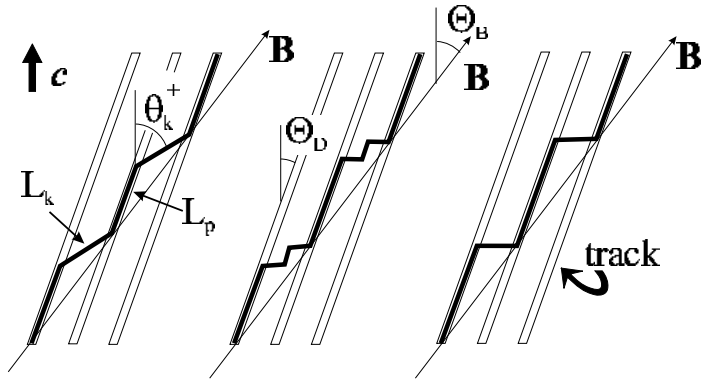}
\caption{Sketch showing the possible vortex configurations when the field is
tilted with respect to the columnar defects.}
\label{staircases}
\end{figure}

Let's now analyze the topology of the vortex staircases. The angle $\Theta
_{k}$ between the kinks and the $c$ axis (see Fig.~\ref{staircases}) can be
calculated by minimization of the free energy\cite{blatter94}. If $L_{p}$ is
the length of a pinned segment, and $L_{k}$ the length of the kink, the line
energy is $E\propto L_{p}\epsilon _{p}\left( \Theta _{D}\right)
+L_{k}\epsilon _{f}\left( \Theta _{k}\right) $, where $\epsilon _{f}\left(
\Theta _{k}\right) \approx \varepsilon _{0}\varepsilon \left( \Theta
_{k}\right) \left[ \ln \kappa +0.5\right] $ and $\epsilon _{p}\left( \Theta
_{D}\right) \approx \varepsilon _{0}\varepsilon \left( \Theta _{D}\right) %
\left[ \ln \kappa +\alpha _{t}\right] $ are the line energy for free and
pinned vortices respectively and $\alpha _{t}<0.5$ parameterizes the core pinning energy
due to the tracks (smaller $\alpha _{t}$ implies stronger pinning).
Minimizing $E$ with respect to $\Theta _{k}$ we obtain the two kink
orientations, $\Theta _{k}^{-}$ for $\Theta _{H}<\Theta _{D}$ and $\Theta
_{k}^{+}$ for $\Theta _{H}>\Theta _{D}$.

%

Since the tracks are tilted, $\left|\Theta _k^{-}\right|$ and $\left|\Theta
_k^{+}\right|$ are different. However, those angles are independent of $%
\Theta_H$. As $\left|\Theta_H-\Theta _D\right| $ increases, $\Theta _k^{\pm
} $ remain constant while $L_p$ decreases and the number of kinks increases,
consequently the pinning energy lowers. This accounts for a decreasing $%
M_{i} $ as $\mathbf{H}$ is tilted away from the tracks. In particular, for $%
\Theta_H=\Theta _k^{\pm }$ vortices become straight ($L_p=0$), thus $\varphi
_T^{\pm }=\left| \Theta _k^{\pm }-\Theta _D\right| $ are the trapping angles
in both directions. In general $\Theta _k^{\pm }$ must be obtained
numerically, but for $\varepsilon \tan \Theta _k\ll 1$ and $\varepsilon \tan
\Theta _D\ll 1$ we obtain

\begin{equation}
\tan \Theta _k^{\pm }\approx \tan \Theta _D\pm \frac 1\varepsilon \sqrt{%
\frac{1-2\alpha _t}{\ln \kappa +0.5}}  \label{kink}
\end{equation}

Eq.(\ref{kink}) adequately describes the main features of the asymmetric
region\cite{evidence,kim} in Fig. \ref{MivsH}, and for $\Theta _D=0$ it coincides with the
usual estimates\cite{nel-vin,blatter94} of $\varphi _T$. There is, however,
an important missing ingredient in the standard description presented above,
namely the existence of twins and Cu-O layers, which are additional sources
of correlated pinning. This raises the possibility that vortices may
simultaneously adjust to more than one of them, forming different types of
staircases (see Fig. \ref{staircases}).

\begin{figure}[htb]
\centering
\includegraphics[angle=0,width=105mm]{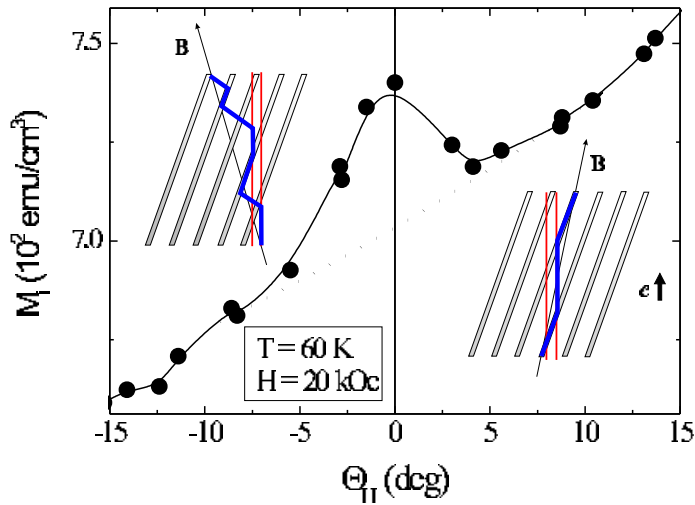}
\caption{Zoom in of the data shown in Fig. \ref{MivsH} for $H=20$kOe around
the $c$ axis. The sketches represent vortex staircases for $\Theta_H>0$ and $%
\Theta_H<0$.}
\label{TBpeak}
\end{figure}

Pinning by twin boundaries is visible in Figure \ref{MivsH} as an additional
peak centered at the $c$ axis for $H=20$ kOe. A zoom in of that peak is shown
in Figure \ref{TBpeak}. The width of this peak, $\sim 10^{\circ }$, is in
the typical range of reported trapping angles for twins\cite
{grigorieva92,fleshler93,li93,oussena96,zhukov97b}. The fact that the peak
is mounted over an inclined background implies that vortices are also
trapped by the tracks. Thus, vortices in this angular range contain segments
both in the tracks and in the twins. These two types of segments are enough
to build up the staircases for $\Theta_H>0$, but for $\Theta_H<0$ a third
group of inclined kinks with $\Theta _k<0$ must exist in order to have
vortices parallel to $\mathbf{H}$ (see sketches in Fig. \ref{TBpeak}).

%

\begin{figure}[htb]
\centering
\includegraphics[angle=0,width=105mm]{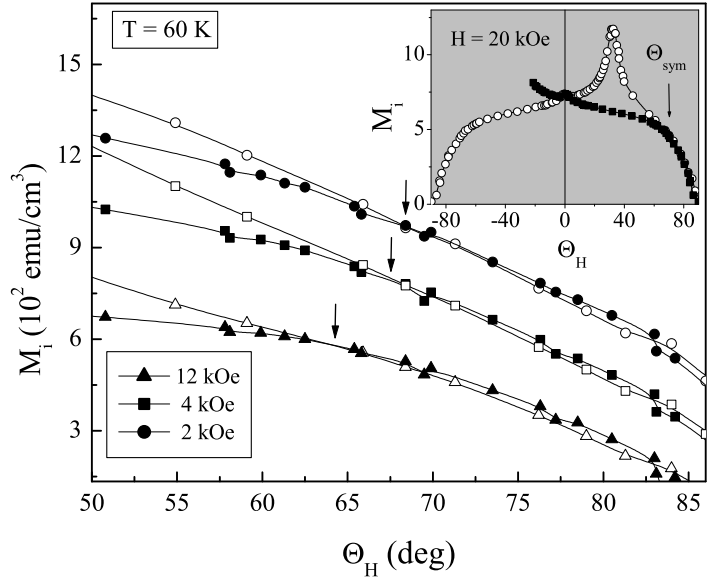}
\caption{Angular dependence of the irreversible magnetization $M_i$ for
three different fields at $T=60$ K. Open symbols: data for $\Theta_H>0$.
Solid symbols: data for $\Theta_H<0$, reflected with respect to the $c$ axis.
Some of the curves were shifted vertically for clarity. The arrows indicate
the angle $\Theta_{sym}$ beyond which the behavior is symmetric with respect
to the $c$ axis. The procedure of reflection of the data is shown in the inset.
}
\label{reflection}
\end{figure}

Another fact to be considered is that the asymmetry in $M_{i}\left(
\Theta_H\right) $ disappears as $\Theta_H$ approaches the $ab$ planes. 
This is illustrated in Figure \ref{reflection}, where $M_{i}$ data
for $-\left| \Theta_H\right| $ was reflected along the $c$ axis and
superimposed to the results for $+\left| \Theta_H\right| $. There is a well
defined angle $\Theta_{sym}$ beyond which $M_{i}\left( \Theta_H\right)$
recovers the symmetry with respect to the $c$ axis (see inset in Figure \ref
{reflection}).

%

\begin{figure}[htb]
\centering
\includegraphics[angle=0,width=105mm]{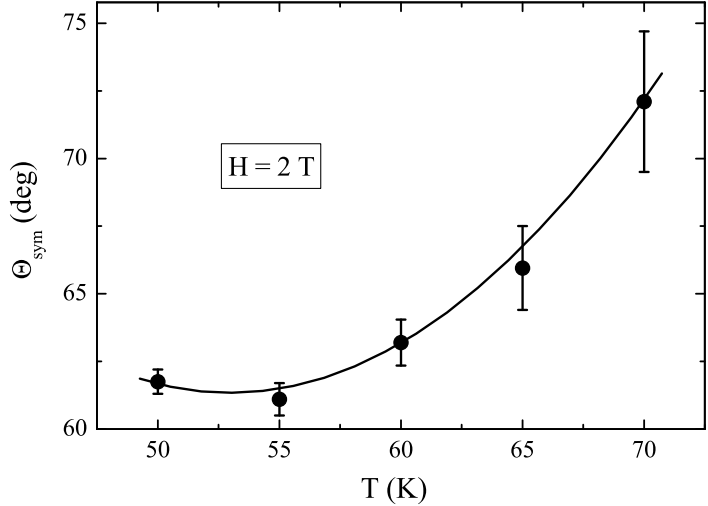}
\caption{Temperature dependence of $\Theta_{sym}$ defined using the
criterium showed in Fig.~\ref{reflection}. The solid line is a guide to the
eye.}
\label{ThetaSym}
\end{figure}

One possible interpretation is that for $\Theta _{H}>\Theta _{sym}$
staircases disappear, \textit{i.e.} $\Theta _{sym}=\Theta _{k}^{+}$, and we are
determining the trapping angle $\varphi _{T}^{+}=\Theta _{sym}-\Theta _{D}$.
However, this is inconsistent with our experimental results. Indeed, $%
\varphi _{T}^{+}$ should decrease with $T$, and this decrease should be
particularly strong above the depinning temperature $T_{dp}\sim 30 - 40$ K due to
the reduction of the pinning energy by entropic smearing effects 
(section I.C). This expectation is in sharp contrast with the observed
increase of $\Theta _{sym}$ with temperature, which is shown in Figure \ref
{ThetaSym} for $H=2$ T. Thus, the interpretation of $\Theta _{sym}$ as a
measure of the trapping angle is ruled out. Moreover, if in a certain
angular range vortices were not forming staircases, pinning could be
described by a scalar disorder strength, then at high fields $M_{i}\left(
\Theta _{H}\right) $ should follow the anisotropy scaling law\cite{blatter92}
$M_{i}\left( H,\Theta _{H}\right) =M_{i}\left( \varepsilon \left( \Theta
_{H}\right) H\right)$. Consistently, we do not observe such scaling in any
angular range\cite{physicaC}.

%

Our alternative interpretation is that, as $\mathbf{H}$ approaches the 
$ab$ planes, the kinks become trapped by the intrinsic pinning. This idea has been used by Hardy {\it et al.}\cite{hardy96} to explain that the $J_c$ at low $T$ in the very anisotropic Bi and Tl compounds with tracks at $\Theta_D=45^{\circ }$ was the same for $H$ either parallel or normal to them. Our situation is somewhat different, as we are comparing two configurations both having kinks.

We first note that, according to Eq. (\ref{kink}), $\Theta _k^{\pm }$ cannot
be exactly $90^{\circ }$ for finite $\varepsilon $, thus the intrinsic
pinning must be incorporated into the model by assigning a lower energy to
kinks exactly parallel to the $ab$ planes. Vortices may now form structures consisting
of segments trapped in the columns connected by segments trapped in the 
$ab$ planes, or alternatively an inclined kink may transform into a
staircase of smaller kinks connecting segments in the planes (see Fig. \ref
{staircases}). We should now compare the energy of the new configurations
with that containing kinks at angles $\Theta _k^{\pm }$. This is equivalent
to figure out whether the kinks at $\Theta _k^{\pm }$ lay within the
trapping regime for the planes or not. The problem with this analysis is
that, as $\Theta _k^{\pm }$ are independent of $\Theta_H$, one of the two
possibilities (either inclined or trapped kinks), will be the most favorable
for all $\Theta_H$. Thus, this picture alone cannot explain the crossover
from an asymmetric to a symmetric regime in $M_{i}\left( \Theta_H\right)$.

The key concept to be considered in this scenario is the dispersion in the
pinning energy of the tracks\cite{nel-vin,blatter94}. The angles $\Theta _k^{\pm }$ depend on the
pinning strength of the adjacent tracks ($\alpha _t$ in Eq. (\ref{kink})),
thus a dispersion in $\alpha _t$ implies a dispersion in $\Theta _k^{\pm }$.
As $\Theta_H$ increases, it becomes larger than the smaller $\Theta _k^{\pm
} $'s (that connect the weaker defects) and the corresponding kinks
disappear. The vortices involved, however, do not become straight, but
remain trapped by stronger pins connected by longer kinks with larger $%
\Theta _k^{\pm }$. This process goes on as $\Theta_H$ grows: the weaker
tracks progressively become ineffective as the ``local'' $\Theta _k$ is
exceeded, and the distribution of $\Theta _k^{\pm }$ shifts towards the 
$ab$ planes. When a particular kink falls within the trapping angle
of the planes, a switch to the pinned-kink structure occurs. In this new
picture, the gradual crossover to the symmetric regime as $\left|
\Theta_H\right|$ increases takes place when most of the remaining kinks are
pinned by the planes.

If kinks become locked, the total length of a vortex that is trapped inside
columnar defects is the total length of a track, independent of $\Theta_H$,
and the total length of the kinks is $\propto \left| \tan \Theta_H - \tan
\Theta _D \right|$. As $\left| \Theta_H\right| $ grows, the relative
difference between the line energy in both orientations decreases, an effect
that is reinforced by the small line energy of the kinks in the $ab$ planes. 
If kinks are not locked but rather form staircases, taking into
account that the trapping angle for the $ab$ planes is small\cite{fleshler93},
the same argument still applies to a good approximation. The temperature
dependence of $\Theta_{sym}$ is now easily explained by a faster decrease of
the pinning of the $ab$ planes with $T$ as compared to the columnar
defects.

Additional evidence in support of our description comes from transport
measurements in the dc-flux transformer configuration where, in the liquid
phase in \textit{twinned} YBCO crystals (without CD) vortices remain correlated along the
$c$ axis \textit{for all field orientations}\cite{morre97}, in contrast to the
observed behavior in untwinned crystals. This suggest that, for all angles,
vortices are composed of segments in the twins and in the $ab$ planes.


\subsection{Influence of the Material Anisotropy and Sample Geometry}


In the previous section we showed that the presence of columnar defects in
HTSC increases the critical current due to the strong pinning and the
reduced vortex wandering when flux lines are trapped into the pinning
potential. We have also shown that at high fields, $J_c(\Theta_H)$ exhibits
a peak when the applied field is aligned with the tracks orientation. As
field decreases, two distinct phenomena progressively and simultaneously
appear. On one hand, we observe a plateau which reflects the existence of a
lock-in phase. On the other hand, we noticed that the peak shifts towards
the $c$ axis. In this section we present a throughout study of the origin of
this displacement and demonstrate that this shift is a consequence of the
misalignment between the external and internal field owing to the
competition between anisotropy and geometry effects.


\subsubsection{Relationship between Internal and Applied Fields}


In thermodynamic equilibrium, the internal field $\mathbf{B}$ is determined
by minimization of the free energy\cite{blatter94} 
\mbox{$G({\bf B})=F({\bf B})-\frac{B^2}{8\pi}+\frac{({\bf B}-{\bf H}){\bf M}}{2}$}, 
where \mbox{${\bf H}={\bf B}-4\pi(1-\hat\nu){\bf M}$}. The
components of the demagnetization tensor $\hat{\nu}$ at the sample principal
axes are $(\nu _{x},\nu _{y},\nu _{z})$, with $\nu _{x}+\nu _{y}+\nu _{z}=1$%
. We adopt the notation that $z$ coincides with the crystallographic $c$ axis,
and that the $x$ axis is perpendicular to both c and $\mathbf{H}$. By
standard minimization of $G(\mathbf{B})$ with respect to $B_{y}$ and $B_{z}$%
, and using the free energy $F(\mathbf{B})$ for the intermediate fields
regime ($H_{c1}\ll H\ll H_{c2}$) we obtain,

\begin{equation}
sin(\Theta_B-\Theta_H)=-\frac{f(\nu_y,\nu_z,\varepsilon)\sin(2\Theta_B)}{%
8\kappa^2} \frac{\ln h+1}{h}  \label{eq:scaling}
\end{equation}
where $f(\nu_y,\nu_z,\varepsilon)=(1-\nu_z)-(1-\nu_y)\varepsilon^2$, ${%
\Theta_B}$ is the direction of the internal field, and the reduced field $h=
H/H_{c2}(\Theta_B,T)$.

The result (\ref{eq:scaling}) only assumes uniaxial anisotropy and the
coincidence of one principal axis with the $c$ axis, and it shows that under
those very general conditions the misalignments due to both mass anisotropy
and sample geometry \textit{have the same field and temperature dependence}.
The function $f(\hat\nu,\varepsilon)$, which contains the combined effects
of geometry and anisotropy, is the key ingredient of the low field behavior,
as its sign determines whether $\Theta_B$ leads or lag behind $\Theta_H$.

To be more specific, let's consider the typical platelike shape of all the
studied single crystals, with thickness $\delta$ along the $c$ axis much
smaller than the lateral dimensions $L_{x}$ and $L_{y}$. To a first
approximation $\nu_{x}= \delta/L_{x}$ and $\nu_{y}=\delta/L_{y}$, 
thus \mbox{$\nu_x, \nu_y, (1-\nu_z)~\ll~1$}. If the material is strongly 
anisotropic and the crystal is not too thin, then \mbox{$(1-\nu_z)~>~(1-\nu_y)\varepsilon^2$}, 
thus $f~>~0$ and \mbox{$\Theta_B~>~\Theta_H$}. We will call this the ``anisotropy-dominated''
situation. In contrast, for thin enough samples of a not too anisotropic
material \mbox{$(1-\nu_z)~<~(1-\nu_y)\varepsilon^2$}, so \mbox{$\Theta_B~<~%
\Theta_H$}. This is what we will call the ``geometry-dominated'' case. The
extreme limit of this case, with an infinite slab ($\nu _{x}=\nu _{y}=0$)
and ignoring the anisotropy, has been discussed by Klein {\it et al.}\cite{klein93}. 
It is also worth to note that for an infinite cylinder with axis
perpendicular to $\mathbf{H}$, where the geometry effects are expected to
cancel out, $\nu _{x}=0$ and \mbox{$\nu_y=\nu_z=\frac {1}{2}$}, thus 
\mbox{$f \propto \left( 1-\varepsilon^2 \right)$} and Eq. (\ref{eq:scaling}) reduces
to the well known expression for the bulk\cite{evidence,blatter94}.

The Eq. (\ref{eq:scaling}) allows us to determine which should be the vortex
direction $\Theta_B$ for a given angle $\Theta_H$ of the controlled variable 
$\mathbf{H}$. In other words, we might use the columnar defects as detectors 
of the internal field orientation taking profit from
the fact that $J_c$ maximizes when $\Theta_B=\Theta_D$. Thus, if we know $%
\Theta_D$ and $\Theta_{max}$, we are able to determine the misalignment $%
\Theta_B-\Theta_H=\Theta_D-\Theta_{max}$. Although the \textit{sign} of such
misalignment is solely determined by the sign of $f$, its \textit{magnitude}
also depends on additional factors such as $\sin(2\Theta_D)$ and $\kappa^2$.
Besides that, the misalignment is strongly temperature and field dependent.
It is easy to see from Eq. (\ref{eq:scaling}) that $\Theta_B \rightarrow
\Theta_H$ for large enough $h$.


At this point it is important to note that, although the misalignment between $\bf{B}$ and $\bf{H}$ is a low field effect, 
Eq.~(\ref{eq:scaling}) can only be used for $H \gg H_{c1}$. It turns out that all our data are well described by eq. (\ref{eq:scaling}). However, the very dilute vortex limit is conceptually interesting, and a detailed discussion about it can be 
found in Ref. \onlinecite{NbSe}.

Table I summarizes the information about geometrical dimensions, mass
anisotropy, dose-equivalent matching field $B_\Phi$ and angle $\Theta_D$ of
the CD with respect to the $c$ axis, for all the crystals that we will refer
in this section.

\begingroup
\squeezetable
\begin{table}[tbh]
\caption{Irradiation and shape specifications for all the crystals studied in 
this section. Crystal B was grown at the T.J. Watson Research Center of IBM 
and irradiated at the Holifield accelerator, Oak Ridge (USA) with 580 MeV
Sn$^{30+}$.}
\centering 
\begin{tabular}[b]{ccccccccccc}
Crystal & material & $\varepsilon ^{-1}$ & $B_{\Phi }(kOe)$ & $\Theta _{D}$
& $\delta (\mu m)$ & $L_{y}(\mu m)$ & $L_{x}(\mu m)$ & $\nu _{y}(\times
10^{-3})$ & $\nu _{x}(\times 10^{-3})$ & $f(\hat{\nu},\varepsilon )(\times
10^{-3})$ \\ \hline
A & $YBa_{2}Cu_{3}O_{7}$ & 7 & 30 & $32^{\circ }$ & 8.5 & 210 & 630 & 40 & 
13.5 & +34 \\ 
B & $YBa_{2}Cu_{3}O_{7}$ & 7 & 30 & $30^{\circ }$ & 20.9 & 715 & 2150 & 29 & 
9.7 & +19 \\ 
C & $YBa_{2}Cu_{3}O_{7}$ & 7 & 57 & $30^{\circ }$ & 11.5 & 1050 & 1050 & 11
& 11 & +1.8 \\ 
D & $YBa_{2}Cu_{3}O_{7}$ & 7 & 22 & $57^{\circ }$ & 4.3 & 381 & 762 & 11.3 & 
5.6 & -3.2 \\ 
E & $NbSe_{2}$ & 3 & 0.5 & $27^{\circ }$ & 7.7 & 765 & 640 & 10.1 & 12 & -67
\\ 
F & $NbSe_{2}$ & 3 & 0.5 & $27^{\circ }$ & 7.7 & 585 & 640 & 13.2 & 12 & 
-63.5 \\ 
G & $NbSe_{2}$ & 3 & 0.5 & $27^{\circ }$ & 7.7 & 419 & 640 & 18.4 & 12 & -58
\end{tabular}
\end{table}
\endgroup

\subsubsection{Anisotropy-dominated case}

\begin{figure}[htb]
\centering
\includegraphics[angle=0,width=150mm]{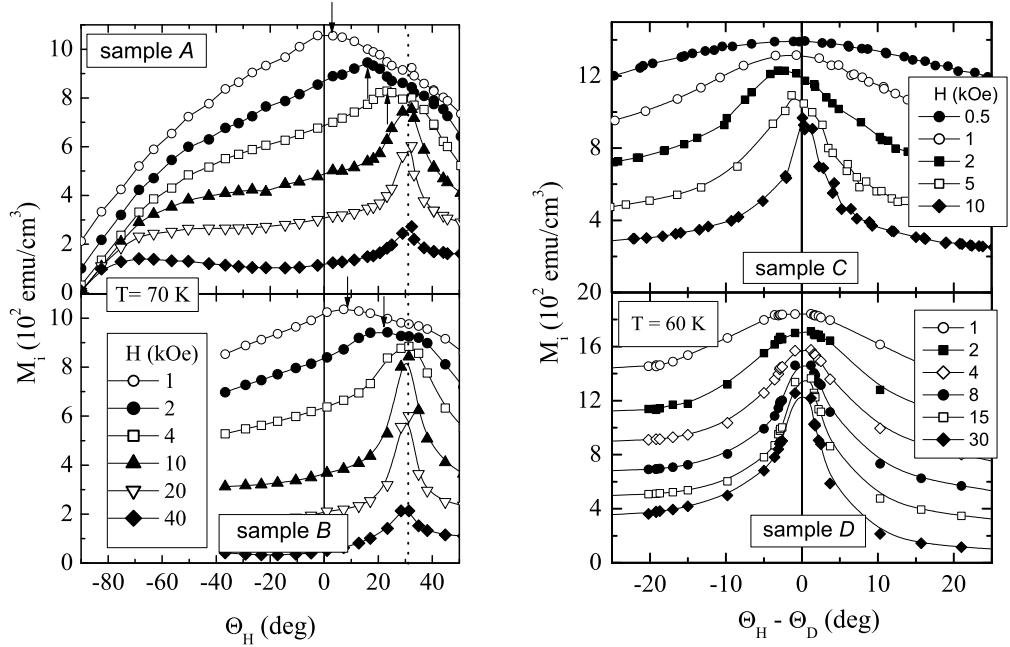}
\caption{{\protect\small Left panel: angular dependence of the irreversible
magnetization for several fields at $T=70$ K for the samples $A$ and $B$
(YBCO). The arrows indicate the angular position, $\Theta_{max}$, of the
maximum in $M_i(\Theta_H)$ for the lowest fields. Right panel: irreversible
magnetization as a function of the relative angle $\Theta_H-\Theta_D$ for
several fields at $T=60$ K in samples $C$ and $D$ (YBCO). For clarity, some
curves have been translated vertically.}}
\label{anisotropy}
\end{figure}

In Fig.~\ref{MivsH} we have already shown the data corresponding to sample $%
A $ where we noticed that the maximum in $J_c$ moves towards the $c$ axis as $H$
decreases. As shown in Fig.~\ref{anisotropy}, this effect persist at $T=70$
K. A similar behavior is observed in crystal $B$ (see Fig.~\ref{anisotropy}%
), although the shift turns out to be smaller than in $A$ at $T=70$ K. These
two crystals have the same anisotropy and irradiation conditions, but
different shapes. Thus, at the same $T$ and $H$ all factors in Eq.~(\ref
{eq:scaling}) are identical, except for $f(\hat\nu,\varepsilon)$. As seen in
Table~I, the difference in demagnetizing factors results in a smaller $%
f(\hat\nu,\varepsilon)$ for sample $B$ than for $A$. Hence, the misalignment
in sample $B$ is expected to be smaller, as indeed observed.

%


\subsubsection{Compensated case}


A striking result predicted by Eq.~(\ref{eq:scaling}) is that the competing
effects (anisotropy and geometry) could be exactly compensated if one were
able to tune the demagnetizing factors and the anisotropy in order to get $%
f(\hat\nu,\varepsilon)=0$, a condition that is satisfied for $%
1-\nu_z=\left(1-\nu_y\right)\varepsilon^2$. For the YBCO single crystals, $%
\varepsilon \sim 1/7$, and this requires extremely thin samples with a big
area. Table I shows that crystals $C$ and $D$ almost satisfy this
compensating condition, as the absolute values of their $f(\hat\nu,%
\varepsilon)$ are, respectively, a factor of $\sim 20$ and $\sim 10$ smaller
than in $A$. The right panel of Fig.~\ref{anisotropy} shows the angular
dependence of $M_i$ for these two crystals at $T=60$ K. Since the CD
orientations $\Theta_D$ in samples $C$ and $D$ are different, in order to
compare them we set the abscise as the relative angle $\Theta_H-\Theta_D$.
In this figure we clearly observe that the peaks remain locked at the tracks
direction even for the lowest fields, in complete agreement with the
expectation. The same behavior was observed for other temperatures.


\subsubsection{Geometry-dominated case}


So far, the samples studied were YBCO crystals with the same anisotropy but
different geometries. On these samples we observed that the peak either
shifts in the direction of the $c$ axis or almost does not deviate from the CD
direction. As pointed out previously, this behavior arises from the strong
anisotropy effect in this material. In order to change the sign of the
deviation (\textit{i.e.}, a shift toward the $ab$ plane), we need to reduce the
anisotropy effects. (Table I shows that crystal $D$ has \mbox{$f<0$}, thus
strictly speaking it is in the geometry-dominated case, but the shift is too
small to be detected).

\begin{figure}[htb]
\centering
\includegraphics[angle=0,width=90mm]{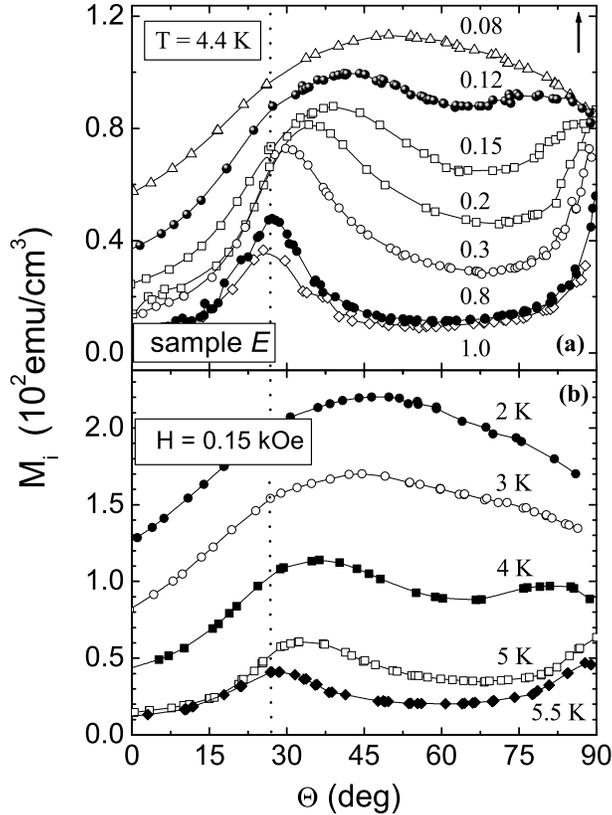}
\caption{{\protect\small Angular dependence of the irreversible
magnetization for sample $E$ (NbSe$_2$) at (a) $T=4.4$ K for several fields
(in units of kOe), (b) $H=0.15$ kOe for several temperatures.}}
\label{geometry}
\end{figure}

To that end we decided to measure NbSe$_2$ single crystals, which have $%
\varepsilon \sim 1/3$, making the anisotropy effect about 5 times smaller
than in YBCO. Beside this, very large and thin NbSe$_2$ crystals can be
readily found, and they can be tailored at will to obtain the desired shape.
Thus, we irradiated a rectangular crystal (labelled as sample $E$), such
that it is in the extreme geometry-dominated case, with $f(\hat\nu,%
\varepsilon) \ll 0$. Fig.~\ref{geometry}(a) shows $M_i(\Theta_H)$ for this
sample at $T=4.4$ K for several $H$. At high fields we observe a peak at the
tracks' direction $\Theta_D=27^{\circ}$. As field decreases the peak becomes
broader and, in contrast to the YBCO behavior, it progressively moves away
from $\Theta_D$ toward the $ab$ plane, in agreement with a negative $%
f(\hat\nu,\varepsilon)$.

%

\begin{figure}[htb]
\centering
\includegraphics[angle=0,width=90mm]{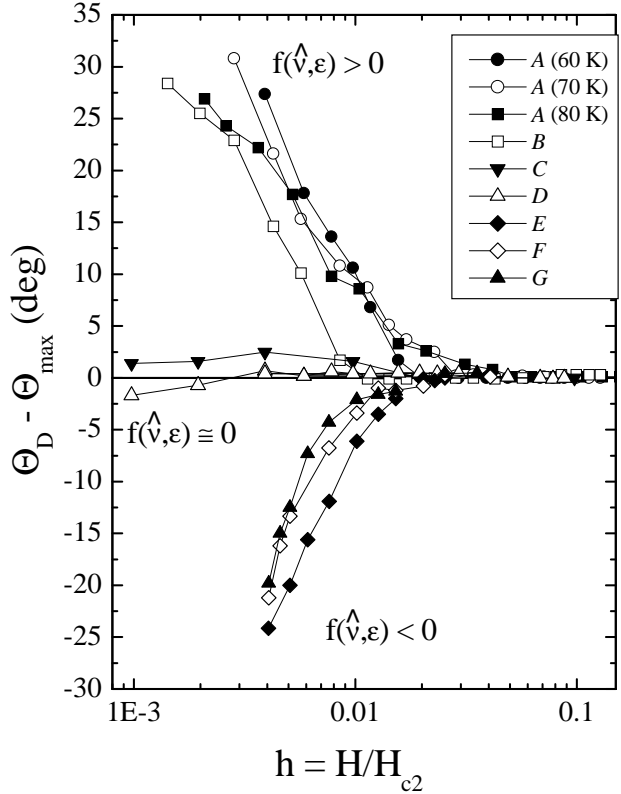}
\caption{{\protect\small Deviation in the maximum of the irreversible
magnetization with respect to the tracks' direction $\Theta_D-\Theta_{max}$,
as a function of $h=H/H_{c2}$, for all samples studied in the present work.}}
\label{Nbpieces}
\end{figure}

The conclusive evidence that the function $f(\hat\nu,\varepsilon)$ dominates
the behavior of the misalignment $\Theta_B-\Theta_H$ comes from samples $F$
and $G$, which are pieces of crystal $E$. These samples were obtained by
cutting the sample $E$ along a line parallel to its shortest side, in such a
way that the demagnetizing factor $\nu_x$ remains unaltered, but $\nu_y$
increases. In this way, the absolute value of the function $%
f(\hat\nu,\varepsilon)$ was progressively reduced, \textit{i.e.}, we moved away from
the ``extreme geometry-dominated case'' and approached the ``compensated
case'' (see Table I). According to Eq.~(\ref{eq:scaling}), the deviation of
the maximum in $M_i(\Theta_H)$ for given $H$ and $T$ should become
progressively smaller for crystals $F$ and $G$ as compared to crystal $E$.
This is in fact observed, as demonstrated in Fig.~\ref{Nbpieces}, where $%
\Theta_{max}-\Theta_D$ for crystals $E$, $F$ and $G$ is plotted as a
function of $h=H/H_{c2}(T,\Theta_H)$.

%

The misalignments for all the YBCO crystals shown in Figs.~\ref{MivsH} and~%
\ref{anisotropy} are also included in Fig.~\ref{Nbpieces}. Thus, this figure
summarizes all the samples studied in the present work, at various
temperatures and fields. The three possible low field behaviors are clearly
visible: anisotropy-dominated (upward curvature), geometry-dominated
(downward curvature) and compensated (almost horizontal curves). It is worth
to note that, in all the not-compensated cases and for both materials, the
misalignment between $\mathbf{B}$ and $\mathbf{H}$ becomes relevant for
fields below a certain characteristic field $H \sim 0.02 H_{c2}$.


\subsubsection{Quantitative test of the model}


The above results clearly demonstrate that the qualitative differences in
the low field behavior are controlled by the factor $f(\hat\nu,\varepsilon)$%
. We now want to verify whether the $H$ and $T$ dependence of the shift is
well described by the model. According to Eq.~(\ref{eq:scaling}) these two
variables appear only through the combination $h= H/H_{c2}(\Theta_B,T)$.
Thus, $\left| \Theta_D-\Theta_{max} \right|$ should increase not only with
decreasing $H$ at fixed $T$, as already seen in figs.~\ref{anisotropy} and 
\ref{geometry}(a), but also with decreasing $T$ at fixed $H$, due to the
increase in $H_{c2}(T)$. This second expectation is also verified, as shown
in Fig.~\ref{geometry}(b) where $M_i(\Theta_H)$ for sample $E$ was plotted
at constant field $H=0.15$ kOe for several temperatures.

\begin{figure}[htb]
\centering
\includegraphics[angle=0,width=105mm]{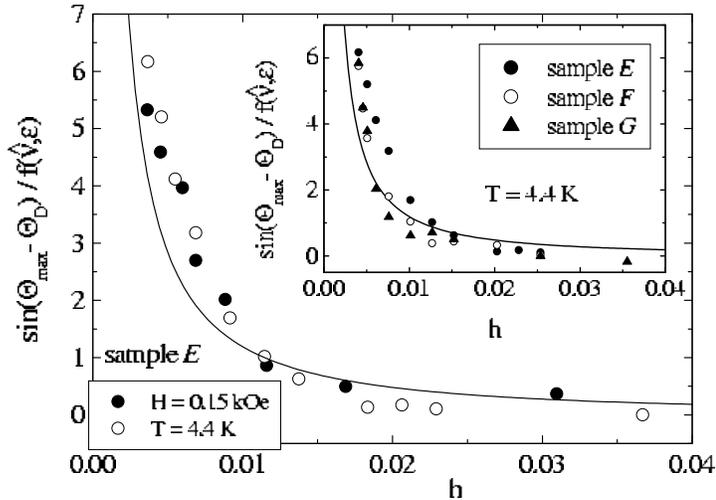}
\caption{{\protect\small Main panel: $\sin(\Theta_{max}-\Theta_D)/f(\hat 
\protect\nu,\protect\varepsilon)$ vs $h$ for the two sets of data shown in
Fig.~\ref{geometry}(a) and \ref{geometry}(b) for the sample $E$. Solid
symbols: $H$ fixed and $T$ swept. Open symbols: $T$ fixed and $H$ swept.
Inset: the same scaling shown in the main panel for samples $E, F$ and $G$.
The solid line in both, main panel and inset, corresponds to Eq. (\ref
{eq:scaling}) with $\Theta_B = \Theta_D = 27^\circ$ and $\protect\kappa = 5.6
$.}}
\label{scaling}
\end{figure}

The equivalence between the variations in $T$ and $H$ is quantitatively
verified in Fig.~\ref{scaling}, where $\sin(\Theta_{max}-\Theta_D)/f(\hat%
\nu,\varepsilon)$ is plotted as a function of $h$ for the two sets of data
shown in Fig.~\ref{geometry}(a) and \ref{geometry}(b). We observe a good scaling, thus
confirming that $h$ is the appropriate variable. The upper critical field
values $H_{c2}=51.8$ kG$\left(1-t\right)$ were taken from the literature 
\cite{detrey}, thus the superposition of the two curves involves no free
parameters.

%

Finally, we analyze the quantitative effect of the factor $%
f(\hat\nu,\epsilon)$. This factor is a constant for a given sample, so it is
the same for all the data in the main panel of Fig. \ref{scaling}. In
contrast, in the inset we show the same scaling procedure for the crystals $%
E, F$ and $G$ at $T=4.4$ K, so now $f(\hat\nu,\epsilon)$ is different for
each sample, while all the other parameters remain identical. We again
obtain a good superposition of the data, although the scaling is poorer than
in the main panel, probably due to the damage produced in the crystal after
each cut process.

The solid line in the main panel of Fig. \ref{scaling} depicts the
expectation of Eq. (\ref{eq:scaling}), with $\Theta_B = \Theta_D = 27^\circ$
as experimentally determined from the location of the maximum at high
fields, and a single fitting parameter $\kappa = 5.6$. The same curve is
shown in the inset. The value $\kappa = 5.6$ is smaller than the accepted
value\cite{detrey} $\kappa \sim 9$, a similar discrepancy to that observed
when studying the lock-in effect.


\subsubsection{The Role of the Columnar Defects}


It is important to keep in mind that the shift in $\Theta_{max}$ at low $h$
is \textit{not} due to the CD. We are only using them as a passive non
perturbative tool to measure the vortex direction in the bulk of the
samples, what is not easy to do by other methods. In fact, the pinning of
the CD is not contained in Eq. (\ref{eq:scaling}), which arises from the
minimization of a free energy, and thus describes a state of thermodynamic
equilibrium. It is obvious, on the other hand, that the uniaxial pinning of
these correlated structures is relevant and should be included in the
analysis. This is usually done by adding to the free energy a term $F_{pin}$
that accounts for the correction to the vortex self-energy due to the CD,
and then comparing the energy of alternative configurations\cite
{hardy96,blatter94}.

This additional contribution depends on the orientation of the vortices, $%
F_{pin}=F_{pin}\left(\Theta_B\right)$, and it is always negative, reflecting
the fact that, for $\Theta_B \neq \Theta_D$, the CD promote the formation of
staircase vortices whose self energy is lower than that of a straight vortex
at the same average orientation. $F_{pin}\left(\Theta_B\right)$ decreases as 
$\Theta_B$ approaches $\Theta_D$ due to the increase of the core trapped
fraction, and it minimizes for that orientation, when the vortex cores are
totally trapped into the tracks\cite{blatter94}. The key point in the
context of the present study, however, is that the incorporation of $%
F_{pin}\left(\Theta_B\right)$ into the scenario does not modify the previous
results, as we show below.

Let's first consider that $\mathbf{H}$ is applied at the angle $\Theta_H =
\Theta_{max}$ such that, in the absence of pinning and according to Eq.~(\ref
{eq:scaling}), the vortices would be at the angle $\Theta_B = \Theta_D$. If
we now ``turn on'' $F_{pin}\left(\Theta_B\right)$, the only effect will be
to deepen the already existing minimum of the free energy at this
orientation, without changing the angle.

Let's now suppose that $\mathbf{H}$ is applied at an angle $\Theta _{H}$
slightly smaller or slightly larger than $\Theta _{max}$. In the absence of
pinning vortices would respectively orient at angles $\Theta _{B}$ slightly
smaller or slightly larger than $\Theta _{D}$, according to Eq.~(\ref
{eq:scaling}). The addition of the term $F_{pin}\left( \Theta _{B}\right) $
will now shift the vortices towards $\Theta _{D}$, that is, a kind of
effective \textit{angular attractive potential} towards the CD orientation
will develop. In particular, for $\left| \Theta _{H}-\Theta _{max}\right|
<\varphi _{L}$, the influence of $F_{pin}\left( \Theta _{B}\right) $ will be
so strong that the system will minimize its free energy by orienting the
vortices exactly along the CD. In section II.B. we have extensively
studied this effect, that manifests in our measurements as a \textit{plateau}
in $M_{i}(\Theta _{H})$ of width $2\varphi _{L}$. Note that the center of
the plateau coincides with $\Theta _{max}$. Thus, although the relation $%
\Theta _{B}$ vs $\Theta _{H}$ will be modified by the CD, the angle $\Theta
_{max}$, experimentally defined as the maximum in $M_{i}(\Theta _{H})$ or as
the center of the plateau where necessary, will still be given by Eq.~(\ref
{eq:scaling}).





\section{Ac response: Dynamics regimes with columnar defects near the
solid-liquid transition.}

In the previous section we have used dc magnetization measurements to
identify and characterize vortex structures at different orientations deep
in the vortex-solid phase, that is, well below the irreversibility line.
What changes in this picture if we increase T and get very close to $%
H_{irr}(T)$? A first issue is that both the vortex characteristic energy $%
\varepsilon _{0}$ and the effective pinning energy $\varepsilon _{p}$
decrease (the last one particularly fast due to the entropic factor), so the
lock-in angle $\varphi _{L}$ tends to zero. As $H_{c2}(T)$ also decreases,
the misalignment effects described in II.C, which are proportional to $%
H_{c2}/H$ (see Eq. (\ref{eq:scaling})), also diminish. Thus, (except perhaps
at extremely low fields) we expect a sharp peak in $J_{c}$ at the angle $%
\Theta _{D}$. A point that is far less obvious is whether staircases will
still form for all orientations, what is their structure, and whether the
various correlated pinning mechanisms will remain coupled.

Of course $J_c(T) \propto \varepsilon_p$ tends to vanish (for any
orientation of $\mathbf{H}$) as we approach $H_{irr}$, thus magnetization
measurements are not sensitive enough to explore this regime. So we will now
turn our attention to the ac susceptibility, which has also been extensively
used to explore the vortex dynamics in HTSC with both correlated an
uncorrelated disorder. Indeed, this is a natural complement of the dc
magnetization studies. The broad range of amplitudes and frequencies of the
ac field easily accessible with this technique allows the exploration of the
dynamics of the vortex system not only in the critical state but in a
variety of other dynamic regimes.

When comparing both types of studies, a key factor to be taken into account
is the characteristic size of the vortex displacements involved. When
measuring magnetization loops, the field changes produce vortex
displacements much larger than the distance between defects. The excitations
explored by creep measurements, on the other hand, are \textit{half loops}, 
\textit{double kinks} and \textit{superkinks} (see section I.C.),
which expand until the vortex (or bundle) is completely removed from the CD.
So all these measurements involve vortex motion through the pinning
landscape at an \textit{inter-valley} scale. In contrast, ac susceptibility
measurements can be carried out at ac field amplitudes so small that pinned
vortices only perform transverse oscillations that are uniform along the
field direction and whose amplitude is only a fraction of the pinning range
of a columnar defect, \textit{i.e.}, \textit{intra-valley} motion. This particularly
interesting situation, which results in a linear response with a very scarce
dissipation, allows to sense the curvature of the pinning potential wells
\cite{campbell1,campbell2}.

In this section we will use ac susceptibility as a tool to first investigate
the angular regimes that exist in the vortex system in the presence of CD in
the vicinity of the solid-liquid boundary, and then to explore the dynamic
behavior of those regimes in a wide range of current densities and excitation 
frequencies. We construct the dynamic phase diagram of the system in the 
temperature vs. ac magnetic field plane and we determine the various crossover 
fields and currents, as well as the parameters that characterize the different 
pinning regimes. To achieve that, we have developed a couple of useful 
procedures for the analysis of the data. It will become apparent that the 
characterization of the dynamic regimes is essential for the correct 
interpretation of the angular dependence in the ac response.

\bigskip


\subsection{Experimental technique}

\bigskip

To measure the complex AC susceptibility $\chi =\chi ^{\prime }+i\chi "$,
crystals were fixed to one coil of a compensated secondary pair rigidly
built up inside a long primary coil that produces an ac field 
$h_{a}e^{i\omega t} \bf{\hat n}$ which is parallel to the $c$ axis and is very homogeneous
in the volume occupied by the sample. The amplitude $h_{a}$ can be varied
from $5 $ mOe to $8$ Oe and the frequency $f=\omega /2\pi $ from $300$ Hz to 
$100$ kHz. A uniform static field $\bf{H_{dc}}$ up to 1300 Oe is added. The AC
coils setup and the sample can be rotated with a precision of $\sim 1^{\circ}$,
allowing to vary the angle $\Theta $ between the $c$ axis and $\bf{H_{dc}}$.
(Notation: At the high temperature, near $T_{BG}$, where these measurements are performed, 
we can consider $\bf{B} \parallel \bf{H}$. Thus, throughout this section we 
will refer simply to the angle $\Theta$ between the DC field and the $c$ axis, $\Theta=\Theta_H=\Theta_B$.
We also refer to the angle $\varphi =\Theta -\Theta _{D}$ between 
$\bf{H_{dc}}$ and the tracks.)

Measurements are performed in all the cases by slowly warming up the sample.
Generally, several curves at various amplitudes $h_{a}$ or at different
angles $\Theta $ were recorded during each temperature sweep. The scan in
angle was always performed in the same direction to avoid backlash problems.
All the curves of $\chi ^{\prime }$ and $\chi "$ at each frequency are
normalized by the same factor, corresponding to a total step $\Delta \chi
^{\prime }=1$ with $H_{dc}=0$. We carefully looked for and confirmed the
absence of thermal decoupling between sample and thermometer, as well as
measurable heating effects due to the excitation current in the primary coil
at high amplitudes.

In each sample, we define the critical temperature $T_{c}$ as the onset of
the AC transition at $H_{dc}=0$ in the linear AC regime (very low $h_{a}$),
and the zero field transition width $\Delta T_{c}$ as the range with nonzero
dissipation in the same condition. The low irradiation doses applied in all
the samples reported in this section, produce no measurable effect either in
the critical temperature or in the transition width.


\subsection{Angular dependence of the ac response.}

In this section we will identify several angular ranges where pinning is
dominated by different mechanisms. With this scope, we will compare the
overall angular response of irradiated and virgin samples. We present
measurements performed on two twinned $YBa_{2}Cu_{3}O_{7}$ single crystals
from the same batch, both having the same $T_{c}=91.8$ K \ and $\Delta
T_{c}\simeq 0.6$ K. One of them (sample A) was irradiated at room
temperature with $291\ MeV\ Au^{27+}$ions with a dose-equivalent matching
field $B_{\phi }\sim 700$ Oe at an angle $\Theta _{D}=$\emph{\ }$30^{o\text{ 
} }$from the $c$ axis, and the other one (sample B) was used as a reference.

%
\begin{figure}[htb]
\centering
\includegraphics[angle=0,width=170mm]{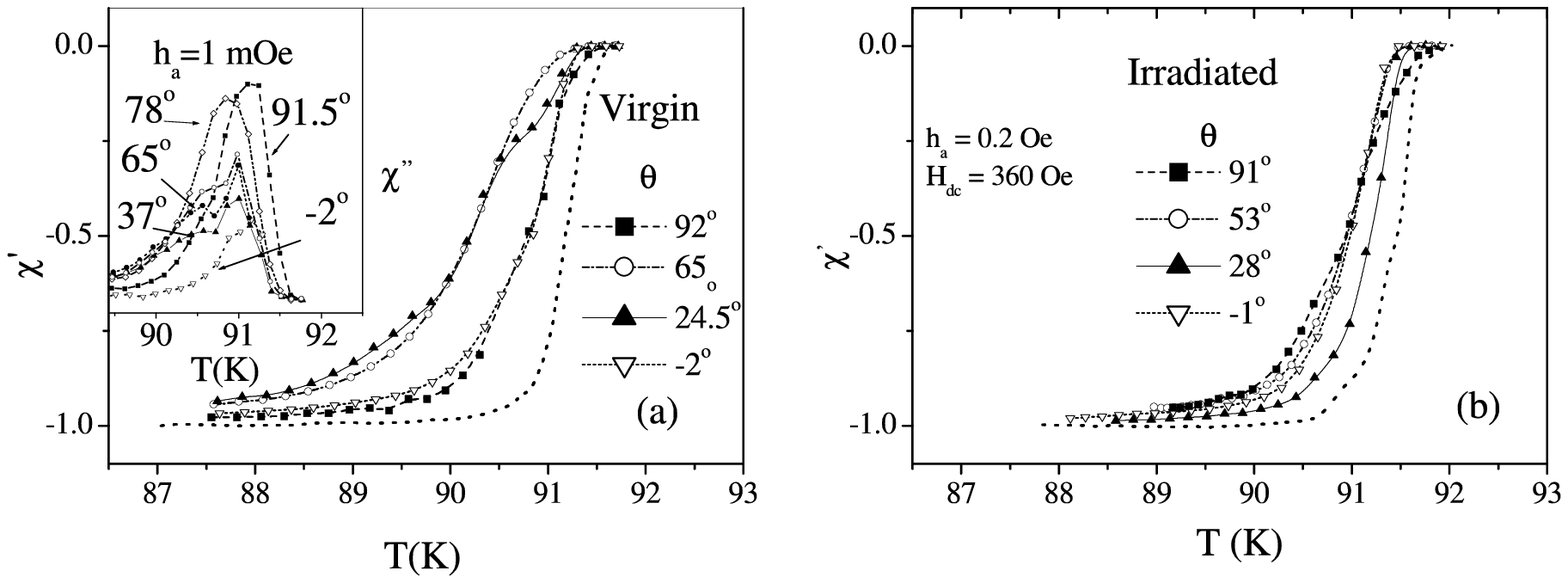}
\caption{{\protect\small a) Temperature dependence of the out-of-phase
component $\protect\chi ^{,}$ of the ac susceptibility with ${\bf H_{dc}}$ 
oriented at various angles $\protect\Theta $ relative to the $c$ axis in a
nonirradiated sample. Inset: curves of $\protect\chi ^{,,}$ as a function of 
$T$ in the linear regime; a structure is present at intermediate angles. b)
Same in a sample irradiated with a matching field $B_{\protect\phi }=700$ Oe
at $\protect\Theta =30^{\circ}$.}}
\label{fig1}
\end{figure}

Figures \ref{fig1}(a) and \ref{fig1}(b) show some of the experimental $\chi
^{,}$ curves recorded as a function of temperature with $H_{dc}=360$ Oe at
different angles $\Theta $ in both the virgin and irradiated crystals. The
zero field transitions are also included as a reference. Comparison of both
figures shows that when the direction of $\bf{H_{dc}}$ is close to the $ab$
planes (curves for $\Theta =92^{\circ} $ in Fig. \ref{fig1}(a) and $\Theta
=91^{\circ}$ in Fig. \ref{fig1}(b), shown with solid squares) the response of
both samples is very similar, indicating that the pinning properties in that
configuration are not significantly altered by the introduction of CD. In
contrast, at all other field orientations the irradiation produces a clearly
visible upward shift of the $\chi ^{,}(T)$ curves, due to the additional
pinning introduced by the CD. Close to the defects ($\Theta =28^{\circ}$, shown
with solid up triangles in the Fig. \ref{fig1}(b)) pinning is drastically
increased.

In the virgin sample, curves at intermediate angles (e.g. $\Theta =24.5^{\circ}$%
, shown with solid up triangles in Fig. \ref{fig1}(a)) present a complicated
structure. This characteristic structure extends to higher $\Theta $ as $%
h_{a}$ decreases, occurring up to around $75^{\circ}$ in the linear regime, as
can be seen in the inset of Fig. \ref{fig1}(b). A similar behavior has been
reported in recent works \cite{sergio}. Generally, the sharp onset is
related with a first order transition and the previous decrease in $\partial
\chi ^{,}/\partial T$ is associated with a peak effect\cite{pico}, although
this last question is still open. Moreover, it has become clear by now that
measurements in nonirradiated crystals at intermediate angles must be done
taking into account the thermomagnetic history\cite{sergio}, so data at 
intermediate angles in Fig. \ref{fig1}(a) must be taken with caution. In any 
case, in the irradiated sample both the structure and memory effects disappear 
completely.

\begin{figure}[htb]
\centering
\includegraphics[angle=0,width=150mm]{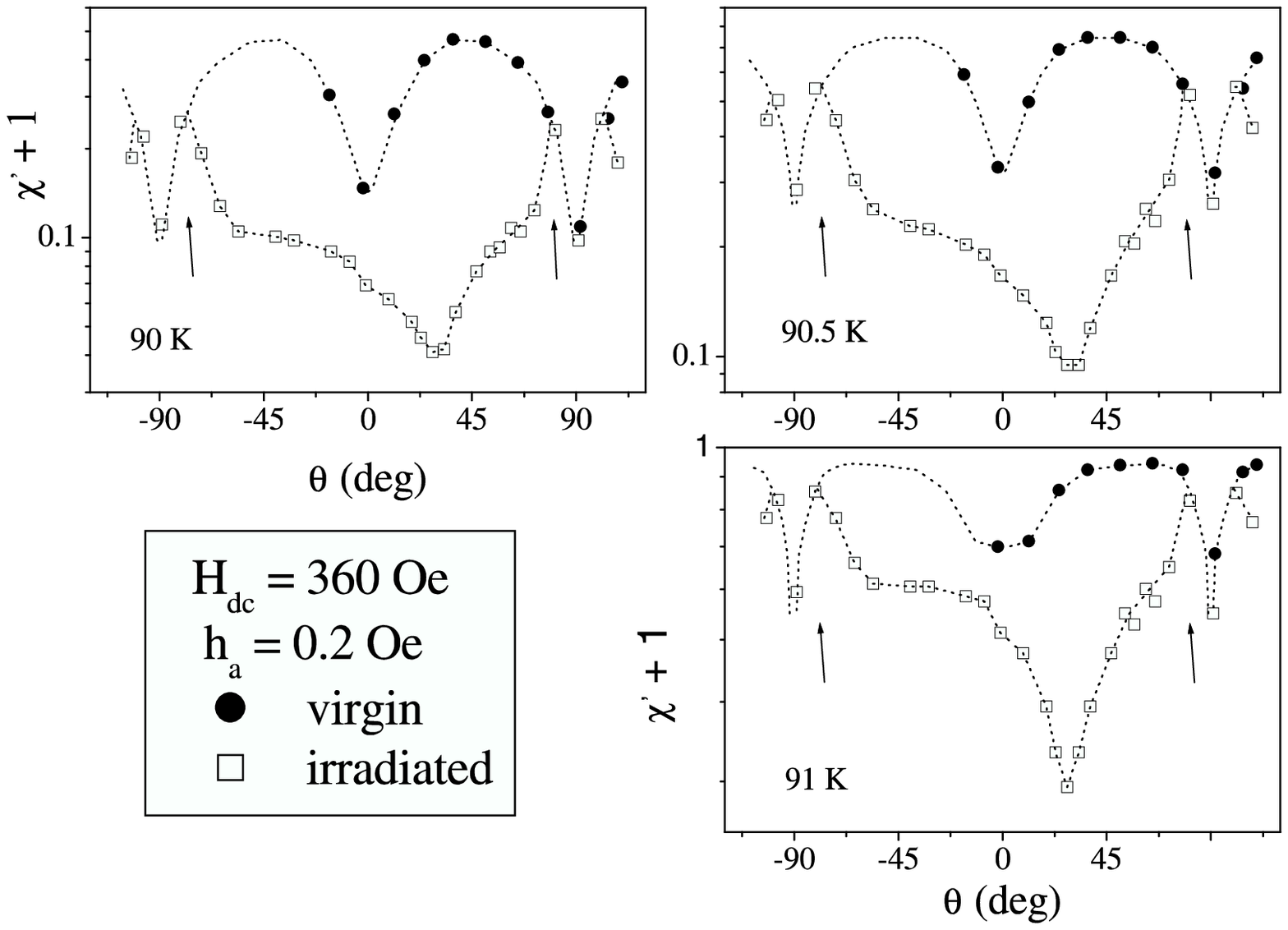}
\caption{{\protect\small Angular dependence of $\protect\chi ^{,}+1$ in an
irradiated (open squares) and virgin sample (circles) at different
temperatures. The dotted lines are guide to the eye (in the virgin sample it
has been constructed from the expected symmetry relative to the $c$ axis). In
the irradiated sample, the symmetry is recovered at $\protect\Theta _{sym} $
(arrows). Beyond this angle responses are very similar.}}
\label{fig2}
\end{figure}

Data as those shown in Fig. \ref{fig1} can be used to build up curves of $%
\chi ^{,}+1$ as a function of $\Theta $ in both samples at fixed $T$. Figure 
\ref{fig2} shows such curves at three different temperatures. Several
features are apparent here. First, in the irradiated sample the original
symmetry is broken ($\chi ^{,}(\Theta )\neq \chi ^{,}(-\Theta )$ ) by the
presence of the tilted defects. This symmetry breaking occurs over a very
large angular range; the symmetry is only recovered for $\left| \Theta
\right| \geq \Theta _{sym}\sim 75^{\circ}$, signaled with arrows in the figure.
Second, in the asymmetric range $\left| \Theta \right| \leq \Theta _{sym}$
the screening of the ac field is, at any temperature, larger than the one
corresponding to the virgin sample. Third, beyond $\left| \Theta
_{sym}\right| $ the behavior of both samples is very similar.

%

The first obvious conclusion that arises from figure \ref{fig2} is that
pinning in the irradiated crystal is dominated by CD over most field
orientations. It is apparent that the CD have a directional effect (\textit{i.e.}
they act as correlated pinning centers), as they modify the original
symmetry. This is true even when $\bf{H_{dc}}$ is almost perpendicular to the
tracks, between $-75^{\circ}$ and $-60^{\circ}$. Note also that, even in that
angular region, the pinning due to the CD is much more efficient than that
present in the non-irradiated sample, as indicated by the much larger
screening (smaller $\chi ^{,})$. In the quadrant of positive $\Theta $
angles (the right side of the figures), where the track direction is
included, the vortex pinning in the irradiated sample abruptly decreases
around the $\Theta _{sym}$ angle, whereas in the left side, the screening
falls down at $\Theta \sim -60^{\circ}$ and, not until $15^{\circ}$ beyond this
angle, at $\Theta \sim -\Theta _{sym}$, the behavior of the non-irradiated
sample is recovered. In section III.E. we will show that the angle
in which the screening falls down (for this case about $-60^{\circ}$) is $h_{a}$
dependent, and is a consequence of a change in the dynamical regime.

Figure \ref{fig2} also shows that the response is not symmetric with respect
to the defects direction. This fact can be easily explained since, as we
argued in section II, the anisotropic character of the material and the
presence of natural correlated defects (twin boundaries and $ab$ planes) should have an
important role in the structure of the vortices.

The main point to emphasize from these measurements is that the angular
region in which the CD act as correlated pinning centers is very large. It
can also be noticed that, in the range of temperature of Fig. \ref{fig2}, $%
\Theta _{sym}$ $\sim 75^{\circ}$ is nearly constant. This behavior is not
consistent with the existence of a trapping angle $\varphi _{T}$
beyond which the correlated nature of the pinning should disappear,
because at these high temperatures a such angle is expected to be very
narrow and to decrease fast with $T$.

We now turn to the situation for field directions close to the $ab$ planes ($%
\left| \Theta \right| >75^{\circ}$). The very similar behavior of both samples
in this angular region suggests that the main pinning source is the same. A
related feature is that, in the non-irradiated sample, a qualitative change
in the linear response is observed very near $\Theta \sim 75^{\circ}$ (inset of
Fig. \ref{fig1}(a)). Beyond this angle, no structure is observed at any AC
field. The above observations could be explained by assuming that $\Theta
_{sym}$ indicates the angle beyond which, for both the irradiated and virgin
samples, the $ab$ planes become the prevailing pinning centers. However, due
to the strong angular dependence of the AC response in the vicinity of the $%
ab$ planes, it is clear that a similar study with a better resolution should
be necessary to confirm the last conclusion. 

It is useful at this point to compare and contrast these results with those obtained by DC magnetization, described in section II. A clear coincidence is that both sets of results show that correlated pinning dominates for all field orientations. The various angular regimes are related to different vortex structures arising from the combined effects of CD, twins and $ab$ planes, not to crossovers to uncorrelated-pinning-dominated regimes. It is also true that, both near the Bose-glass transition and deep into the solid phase, over most field orientations vortices form staircases with segments pinned into the CD. The comparison of the $\Theta_{sym}$ obtained by both techniques is particularly interesting. The similarity of the values (see figs. \ref{reflection}, \ref{ThetaSym} and \ref{fig2}) strongly suggests that the underlying physics is similar. On the other hand, we should look carefully at the meaning of $\Theta_{sym}$ in each case. In the dc magnetization studies, our interpretation is that for $\Theta > \Theta_{sym}$ most of the kinks (which connect the vortex core segments pinned by the CD) are locked into the $ab$ planes. This implies that intrinsic pinning must be important in that angular range. The ac data on Fig. \ref{fig2} additionally suggest that at that high temperature intrinsic pinning may be {\it the dominant} pinning source in that range, as the ac response of the crystals with and without CD is the same. As this is not so clear at lower temperatures, more detailed studies of the interplay of CD and intrinsic pinning near $\Theta_{sym}$ as a function of temperature should be performed. Finally, it is worth to mention that, as expected, no evidence of lock-in phase or misalignment between $\bf B$ and $\bf {H_{dc}}$ is observed at the high temperatures of the ac studies.

Once the overall angular dependence of the ac response has been determined,
we can use the ac susceptibility measurements to explore in detail the
various dynamic regimes that occur at a given field orientation as a
function of $T$ and $H_{dc}$, so a more microscopic information of the
vortex-defects interactions can be obtained. Basically, this is achieved by
changing the amplitude of the vortex oscillations, which are determined by
the amplitude $h_{a}$. By performing these studies at a few selected
orientations we can identify differences among the various angular regions,
as will be shown in the next two subsections.

\bigskip

\bigskip


\subsection{Dynamic regimes of vortices pinned by aligned columnar defects.}

\bigskip

In this section we concentrate in the ac dynamic response when vortices are
aligned to the columnar defects (\textit{i.e.} $\varphi =0$). By varying the
amplitude and frequency of the ac field, we are able to cover a wide range
of current densities. At the lowest amplitudes we observe the linear
response with very low dissipation, characteristic of the oscillation of
pinned vortices inside the tracks (Campbell regime)\cite{campbell1,campbell2}, while
at the high $J$ limit a critical state develops. In between, a large
non-linear transition regime is observed. We explore both the intra- and
inter-valley dynamics in the same experiment, and we construct the dynamic
phase diagram of the system in the $T$ vs. $h_{a}$ plane. We will present
results for two irradiated samples. One of them is the sample A used in the
previous section. The other one (sample C), was irradiated with $280\ MeV\
Sb^{22+}$ at an angle of $15^{\circ}$ from the $c$ axis to a dose equivalent to $%
B_{\phi }\sim 3000$ Oe.


\subsubsection{Linear response}

A \textit{linear regime} is characterized by a $\chi $ independent of $h_{a}$%
. By analyzing both components $\chi ^{,}$ and $\chi ^{,,}$ we are able to
establish the nature of the regime. The linear ac susceptibility $\chi $ due
to vortex motion in a superconductor is determined\cite{Vander2,Coffey1,Brandt1,brandtA}
by the complex penetration depth $\lambda _{AC}(T,B,\omega )=\lambda _{R}+i\lambda _{I}$. 
The ratio $\varepsilon =\lambda_{I}/\lambda _{R}$, identifies several linear regimes. 
An ohmic response (\textit{i.e.} a real resistivity) such as flux flow or thermally assisted
flux flow (TAFF) corresponds to $\varepsilon =1$. A \textit{real} $\lambda _{AC}$ ($\varepsilon =0$)
represents a non-dissipative response with $\chi ^{,,}=0$, which is almost
the case in the Campbell regime where \textit{pinned} vortices perform 
\textit{intra-valley} oscillations, although there is always some residual
dissipation, thus $\varepsilon \ll 1$ but finite.

\begin{figure}[htb]
\centering
\includegraphics[angle=0,width=100mm]{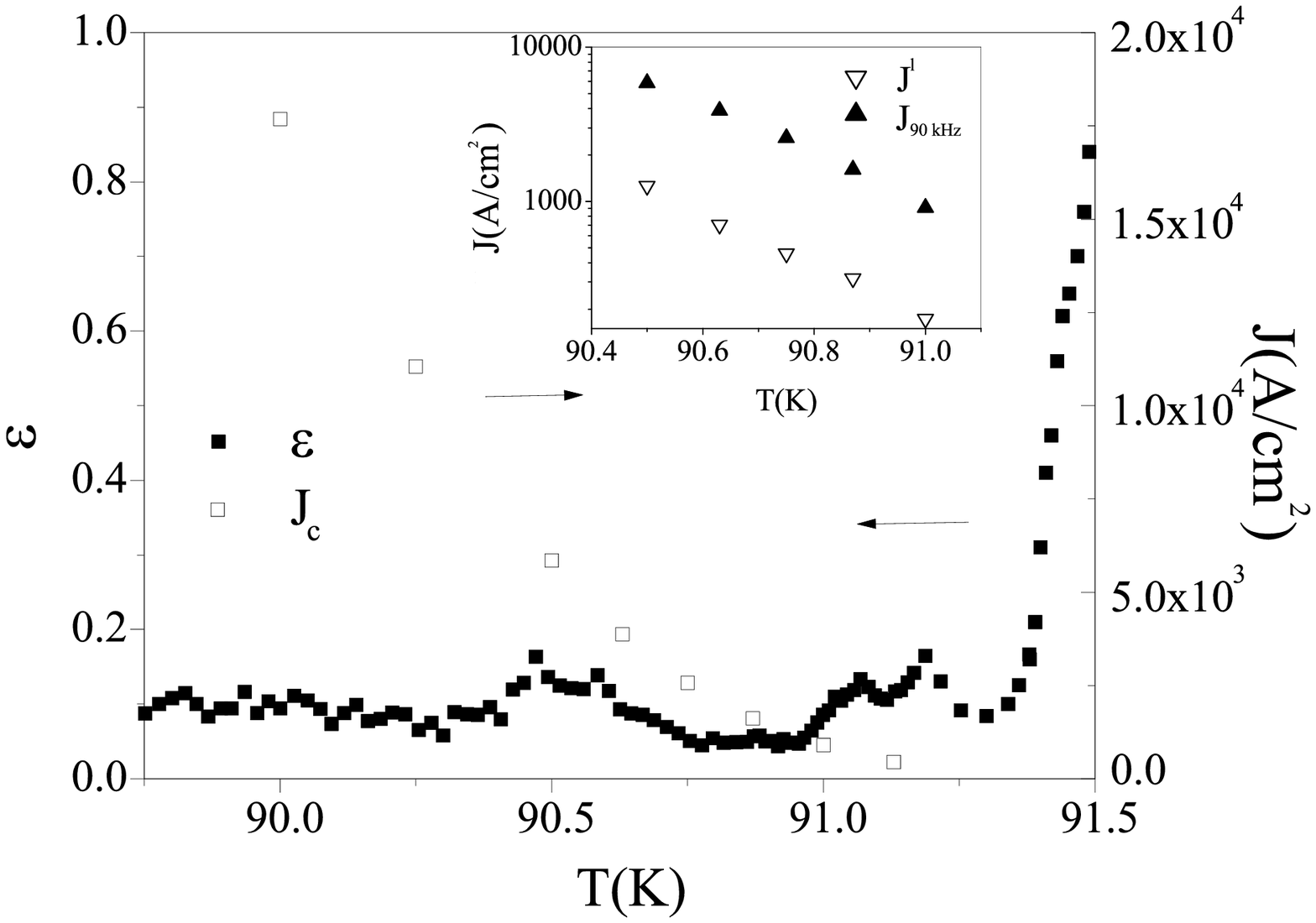}
\caption{{\protect\small Temperature dependence of $\protect\varepsilon $=$%
\protect\lambda _{I}$/$\protect\lambda _{R}$, where $\protect\lambda _{ac}$=i%
$\protect\lambda _{I}$+$\protect\lambda _{R}$ is the penetration depth in
the linear regime, and J$_{90kHz}$ is the frequency dependent current
density in the Bean regime for f=90 kHz. At T=91.3 K there is a sudden
increase in $\protect\varepsilon $ towards the ohmic regime and J$_{90kHz}$
vanishes. The inset shows in semi logarithmic scale the temperature
dependence of J$_{90kHz}$ and J$^{l}$, the limiting current of the linear
regime.}}
\label{fig4}
\end{figure}

Our experimental situation is well approximated by a thin disk in a
transverse ac magnetic field where, according to Brandt \cite{brandtA},

\begin{equation}
\chi ^{,}+1+i\chi ^{,,}=\sum\limits_{n=1}^{N}c_{n}/(\Lambda _{n}+\varphi )
\label{eqbrandt}
\end{equation}
Here $\varphi =R\delta /2\pi \lambda _{AC}^{2}$, where $R$ and $\delta $ are
the disk radius and thickness respectively, $c_{n}$ and $\Lambda _{n}$ are
tabulated real constants, and the sum arises from the discretization
involved in the numerical procedure. \ Expression \ref{eqbrandt} allows us
to calculate $\chi ^{,}(\lambda _{R},\varepsilon )$ and $\chi ^{,,}(\lambda
_{R},\varepsilon )$. We can numerically invert this function to compute $%
\lambda _{R}(T)$ and $\varepsilon (T)$ from our $\chi (T)$ data in the
linear regime \cite{Gabi1}. Figure \ref{fig4} shows $\varepsilon (T)$ so
obtained in sample A for $\bf{H_{dc}}=360$ Oe parallel to the tracks. We observe
that $\varepsilon (T)$ remains small and approximately constant, $%
\varepsilon \sim 0.1$, for $T\leq 91.3K$. This behavior is characteristic of
trapped vortices oscillating inside their pinning wells. At $T\sim 91.3$ K
an abrupt depinning occurs, as indicated by the sudden increase in $%
\varepsilon $ which in a narrow temperature range of $\sim 0.2$ K grows
close to $\varepsilon =1$ that corresponds to the ohmic (flux flow) regime.

%

It can be seen from Eq. (\ref{eqbrandt}) that, when $\varepsilon \ll 1$
(low dissipation limit), $\chi ^{,}$ is solely determined by the real
part of the penetration depth, $\lambda _{R}$. On the other hand, 
$\lambda _{R}^{2}(T,B)=\lambda_{L}^{2}(T)+\lambda _{C}^{2}(T,B)$, where 
$\lambda _{L}$ and $\lambda _{C}$ are the London and Campbell penetration
depths respectively\cite{Vander2,Coffey1}, being

\begin{equation}
\lambda _{C}^{2}(T,B)=\frac{B\Phi _{0}}{4\pi \alpha _{L}}  \label{campbell}
\end{equation}
where $\alpha _{L}$ is the Labusch parameter that measures the restoring
force of the pinning well.

\begin{figure}[htb]
\centering
\includegraphics[angle=0,width=150mm]{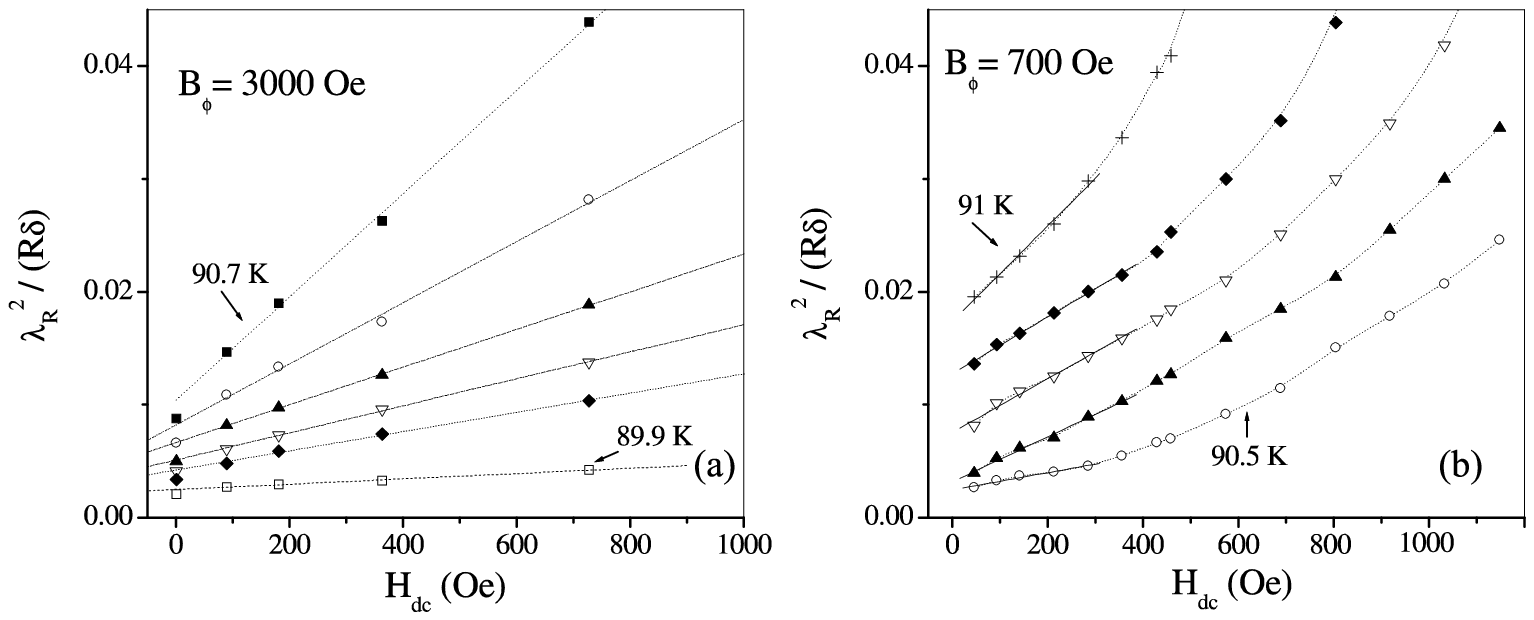}
\caption{{\protect\small Field dependence of the square of the dimensionless
linear real penetration depth at various temperatures in two irradiated
samples with different doses. If $B\lesssim B_{\protect\phi }/2$, a linear
dependence holds.}}
\label{fig5}
\end{figure}

Thus, measurements of $\chi (T)$ in the Campbell regime at various $H_{dc}$
allows us to obtain $\lambda _{C}(T,B\sim H_{dc})$ and $\alpha _{L}(T,B\sim
H_{dc})$. Figure \ref{fig5} shows $\lambda _{R}^{2}$ vs. $H_{dc}\sim B$ at
various temperatures in both samples. In sample C (figure \ref{fig5}(a)), 
$B_{\phi }\sim 360$ Oe without increasing the vortex density. It can be seen
from figure \ref{fig5}(a) that, in sample C, $\lambda _{C}^{2}\propto
H_{dc}\approx B$,\ so Eq. (\ref{campbell}) implies a field-independent $\alpha
_{L}$. This important result strongly suggests that vortices are
individually pinned. The last interpretation is well supported by results in
the sample A (figure \ref{fig5}(b)): at low dc fields $\lambda _{R}^{2}$
grows linearly with $H_{dc}$, but above $H_{dc}\sim 400$ Oe $\sim \frac{1}{2}%
B_{\phi }$ there is a faster increase of $\lambda _{R}^{2}$ with $H_{dc}$,
indicating that $\alpha _{L}$ decreases with $B$. We interpret this feature
as the crossover to a collective Campbell response. The values of $\alpha
_{L}$ obtained in both samples are very similar, demonstrating that the
field-independent $\alpha _{L}$ indeed characterize the elementary vortex-
tracks interaction in a single vortex pinning regime.

%

We also detect a small dissipative component, frequency independent within
our resolution, and which is likely to originate in thermally induced jumps
between two metastable states of similar energy (two level systems) that
characterizes a \textit{glassy phase}\cite{Koshe1}. We will not extend on this
topic here, but refer to Ref. \onlinecite{Gabi1} for more details.


\subsubsection{The onset of nonlinearity}

As $h_{a}$ increases, the response becomes nonlinear. In crystals with
aligned CD the crossover from linear to nonlinear response is very smooth,
so the exact boundary $h_{a}^{l}(T)$ between both regimes is difficult to
determine and consequently we adopted an experimental criterion explained in
ref. \onlinecite{Gabi2}. This boundary signals the appearance of vortex
oscillations that are large enough to break the linear approximation in the
vortex motion equation. This may originate either from intra-valley motion
(breakdown of the parabolic approximation for the pinning potential at large
oscillation amplitudes\cite{campbell1,campbell2}) or from the increase of nonlinear
flux creep (inter-valley jumps) as $J$ grows\cite{Vander2}. In either case,
the decrease of the pinning energy as $T$ increases implies that $h_{a}^{l}$
should decrease with $T$, as indeed observed.

As the amplitude of vortex oscillations in the linear limit is proportional 
\cite{Vander2} to the local current density $J$, which is spatially
inhomogeneous, the breakdown of the harmonic oscillations occur at different 
$h_{a}$ in different parts of the sample. For a disk in a transverse field,
the current distribution in the linear regime can be numerically calculated
starting from the experimental values of $\chi ^{,}(T)$ and following the
procedure described by Brandt \cite{brandtA}. The method allows us to
estimate the actual current density $J(T,\rho ,h_{a})$, where $\rho $ is the
distance from the center of the disk. Since $J$ is larger at the disk border
($\rho =R$), nonlinearities first occur there, when the local current
density reaches the value $J^{l}(T)=J(T,R,h_{a}^{l}(T))$. Using our
experimental estimate of the crossover field $h_{a}^{l}(T)$, we can
calculate $J^{l}(T)$ as shown in the inset of Figure \ref{fig4}. As $h_{a}$
keeps growing the boundary of the Campbell regime moves inward into the disk.

We can also estimate the transverse displacement of the vortices near the edge of
the sample (those that perform the largest oscillations), $u(\rho \sim R)$.
In the linear regime, $u$ is related with $J$ in the same position through
the simple form $(1/c)\,\phi _{0\,\;}J(\rho)\approx \alpha _{L}u(\rho )$.
The vortex displacement at the edge of the sample when linearity breaks
down, $r_{l}$, will be

\begin{equation}
r_{l}=\frac{1}{c}\frac{\phi _{0}}{\alpha _{L}(T)}J^{l}(T)  \label{radiolim}
\end{equation}

The radius $r_{l}$ measures the range where the pinning potential 
$\varepsilon _{p}\left( r\right)$ can be well approximated by a parabola. 
Using values for $J^{l}(T)$ and $\alpha _{L}(T)$ calculated from the
experimental data, we obtain $r_{l}$ $\approx 50$ {\AA} in the sample A 
(irradiated with $Au$ ions), and $r_{l}$ $\approx 25$ {\AA} in the sample B 
(irradiated with $Sn$ ions). These values have a striking coincidence with 
the observed radius of columnar defects $r_{D}$ in each case\cite{marwick}. 
It is somehow surprising that even at these high temperatures where $\xi >r_{D}$, 
and so no drastic change in $\varepsilon_{p}\left( r\right)$ is expected\cite{blatter94} 
at $r=r_{D}$, the method still provide us information on the real size of the tracks.


\subsubsection{The Critical state}

\bigskip

At the highest amplitudes, vortices perform large excursions outside the
pinning wells and inter-valley motion will dominate over most of the sample.
Above a lower boundary $h_{a}^{c}(T)$ the response is well described by a
Bean critical state \cite{bean}. The large intermediate region 
$h_{a}^{l}(T)<h_{a}<h_{a}^{c}(T)$ corresponds to a crossover regime where
both the intra-valley and inter-valley vortex motion contribute
significantly to the ac response.

To define the lower boundary $h_{a}^{c}(T)$ of the critical state regime, we
have developed an experimental method based in the analysis of the $\chi
^{,}(T,h_{a})$ data\cite{Gabi2}. We will not explain the method in detail
here but only point out its basic concepts. The Bean Model assumes that
there is an uniform persistent density current in any region of the sample
where inter-valley vortex motion occurs. Due to flux creep effects, that
current density $J_{\omega }$ is smaller than $J_{c}$ and depends on the
frequency of the measurement. This determines a frequency dependent Bean
length 

\begin{equation}
\Lambda _{c}(T,h_{a})=\frac{c}{4\pi }\frac{h_{a}}{J_{\omega }(T)}
\label{lambdaBean}
\end{equation}

The screening component $\chi ^{,}$ is only a function of $\Lambda _{c}$
divided by some characteristic sample dimension. The dependence of $\chi ^{,}
$ on this dimensionless variable is shape-dependent and in general difficult
to calculate. The key idea of the method is to identify data (at fixed
frequency) at different $T$ and $h_{a}$ that combine to produce the same $%
\Lambda _{c}$ according to Eq. (\ref{lambdaBean}). The consistency of the
resulting function $\Lambda _{c}(h_{a})$\ is checked at each temperature. A
critical state is established if, at constant $T$ and $\omega $, $\Lambda
_{c}$ turns out to be proportional to $h_{a}$.

\begin{figure}[htb]
\centering
\includegraphics[angle=0,width=150mm]{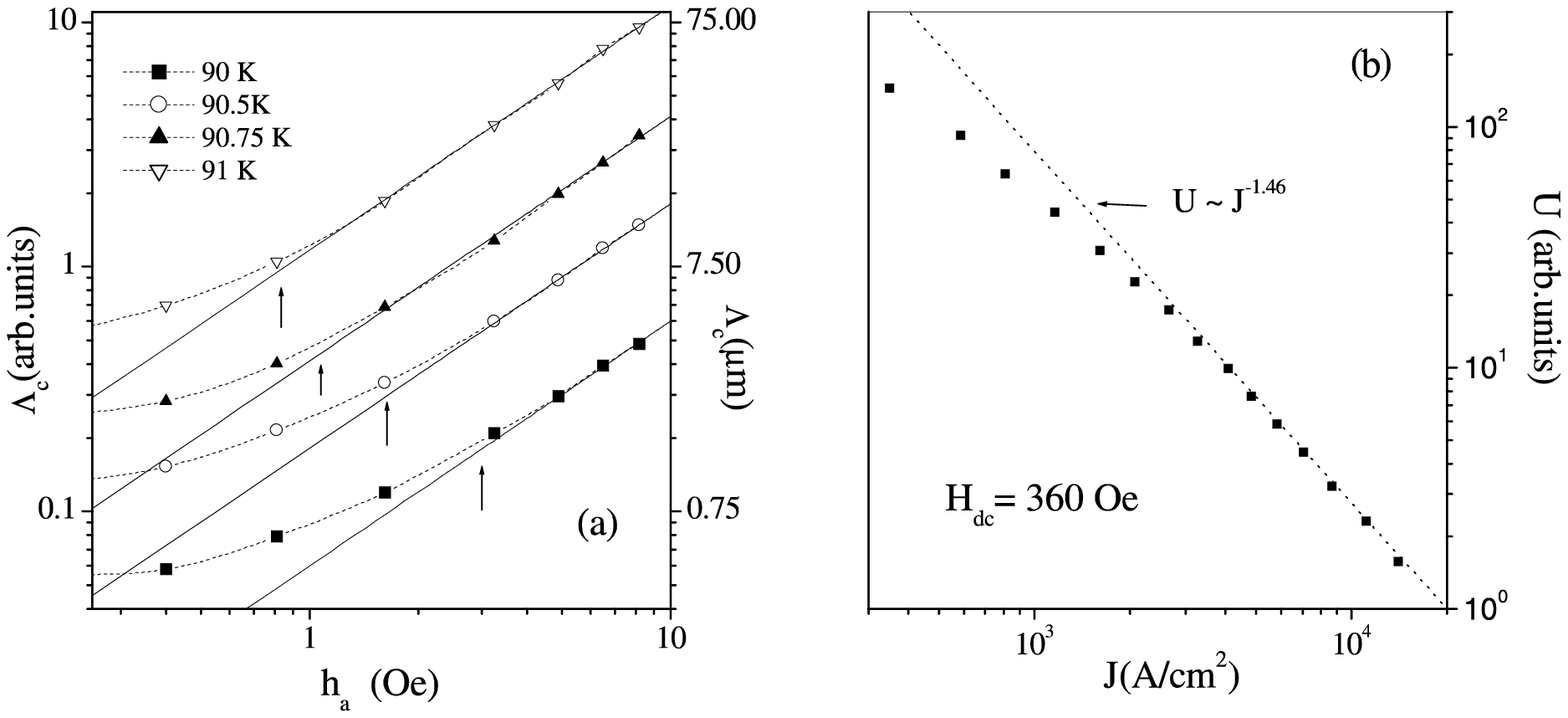}
\caption{{\protect\small (a) Test that proves the existence of a critical
regime. The Bean length $\Lambda _{c}$ is proportional to $h_{a}$ above $%
h_{a}^{c}(T)$ signaled by the arrows. (b) Current density dependence of the
activation energy. The U values came from the frequency dependence of $%
\protect\chi ^{,}$ in the critical state.}}
\label{fig6}
\end{figure}

Figure \ref{fig6}(a) shows the result of such analysis at several
temperatures in sample A. In all cases we observe that $\Lambda _{c}\propto
h_{a}$ at high amplitudes, thus proving the existence of a Bean regime. At
low amplitudes there is a systematic deviation from a straight line and $%
\Lambda _{c}$ becomes larger than expected in the Bean regime, indicating
the absence of a fully developed critical state. The field $h_{a}^{c}(T)$
that marks the onset of the Bean regime, is shown by arrows in Figure \ref
{fig6}(a). Note that in the critical state regime $d\Lambda
_{c}/dh_{a}\propto 1/J_{\omega }(T)$, thus the variation of the slope of $%
\Lambda _{c}$ vs. $h_{a}$ with $T$ (which appears as a vertical shift in the
log-log plot of Figure \ref{fig6}(a)) allows us to determine the \textit{%
temperature dependence} of $J_{\omega }(T)$.

%

The great advantage of the above procedure is that it allows us to test the
existence of a Bean regime, to determine its limits in the $h_{a}-T$ plane
and to obtain the temperature dependence of $J(T)$ \textit{regardless of the
sample geometry}. The function $\chi ^{,}(\Lambda _{c})$ is unique to each
sample, although, as expected, we found\cite{Gabi2} that the general shape is similar
for different YBCO crystals with similar aspect ratio.

To estimate the absolute values of $J$ we need to estimate the absolute
values of $\Lambda _{c}$, and to that end we must rely on some geometrical
modeling. We use the result\cite{clem-san} for a thin disk in a transverse field, $\chi
^{,}=0.5$ when $\Lambda _{c}/\delta =0.75$. In the right
axis of Figure \ref{fig6}(a) we indicate the actual values of $\Lambda _{c}$
obtained in adopting such criterion. The corresponding values of the
persistent current density (identified as $J_{90kHz}\left( T\right) $ to
emphasize that it is frequency dependent) are shown in Figure \ref{fig4}.
The numerical estimates of $\Lambda _{c}\left( T\right) $ and $J\left(
T\right) $ may be affected by a systematic error because our sample is not a
disk, but the difference will be an overall factor of the order of unity and
is irrelevant within the context of our analysis.

Inspection of Figure \ref{fig4} shows that the persistent current vanishes
very close to the temperature where the Campbell regime disappears and $%
\varepsilon $ in the linear limit starts to grow. The coincidence of both
facts identify the \textit{Bose glass} transition temperature $T_{BG}$. In
the inset of figure \ref{fig4}, we can see that the temperature dependence
of the persistent current density $J_{90kHz}$ in the critical state closely
follows that of $J^{l}\left( T\right) $, the limiting current of the linear
regime. This observation implies that the linearity of the intra-valley
oscillations in the Campbell regime is lost when the current density at the
sample perimeter reaches a significant fraction ($\sim 20\%$) of the current
density that flows in the critical state.

As mentioned above, the dynamics in this limit is strongly affected by flux
creep, thus the persistent current density $J$ is much smaller than the
critical current density $J_{c}$ . Measurements at different frequencies
allow us to determine the temperature and current density dependence of the
activation energy, $U(J,T)$ in a wide range of $J$. We have devised\cite{Gabi2} 
a procedure to extract this information from the frequency
dependence of the $\chi (T)$ curves in the critical state. Figure \ref{fig6}%
(b) shows the main result from this study. There is a characteristic glassy
behavior $U(J)=J^{-\alpha }$ (see section I.C.). At large current
densities $U\propto J^{-1.5}$, the exponent progressively decreasing at low
$J$. A dynamic exponent $\alpha =1.5$ is  typical of
collective creep of vortex bundles\cite{blatter94,konczy2}.  Blatter {\it et al.} 
\cite{blatter94} predict this dependence  in the regime of large bundles,
\textit{i.e.}, for transverse bundle dimension larger than $\lambda _{L}$.
(The decrease of $\alpha $ at low $J$ could 
indicate a crossover to a charge density wave (CDW)-type creep regime).
We also observe a very strong $T$ dependence\cite{Gabi2}, consistent with
creep of vortex bundles at high temperature.


\subsubsection{Dynamic diagram}

\begin{figure}[htb]
\centering
\includegraphics[angle=0,width=100mm]{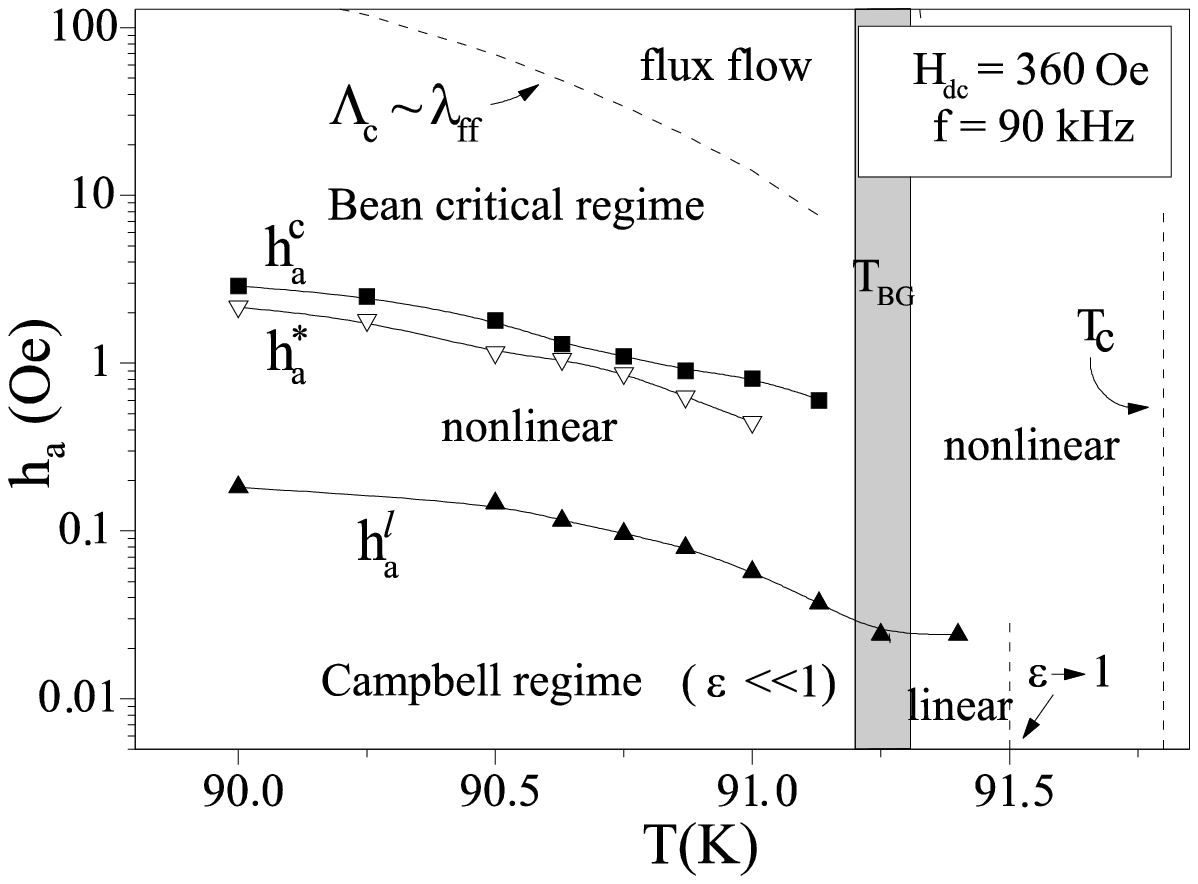}
\caption{{\protect\small Dynamic diagram in the $(h_{a},T)$ plane for an
irradiated sample with $B_{\protect\phi }=700$ Oe.}}
\label{fig7}
\end{figure}

From the above analysis, it is possible to build up a diagram
in the $(h_{a},T)$ plane to identify the various dynamic regimes and analyze
the crossovers among them. Figure \ref{fig7} shows a such diagram, for the
sample A at $H_{dc}=360$ Oe parallel to the CD, and $f=90$ kHz. The response
for $T\leq 91.2$ K is characteristic of pinned vortices in a glassy phase.
Below the crossover field $h_{a}^{l}(T)$, we observe a regime of pinned
vortices performing intra-valley oscillations inside their pinning sites.
The ac susceptibility is well characterized by a Campbell penetration depth
and a small dissipative component. Above a second line $h_{a}^{c}(T)$ (and
again for $T\leq 91.2$ K), a Bean critical state forms. The persistent
current density flowing in this regime is determined by thermally activated
processes that involve inter-valley collective jumps of vortex bundles. In
between these two extreme regimes there is an intermediate nonlinear region
that occupies a large portion of the $(h_{a},T)$ plane, as $h_{a}^{l}(T)$
and $h_{a}^{c}(T)$ differ by more than one order of magnitude as seen in the
Figure.

%

Within the temperature range $T\sim 91.2-91.3$ K both the Campbell and Bean
critical regimes disappear. This is indicative of a sudden depinning of
vortices at a Bose-glass transition that occurs in the range $91.2$ K $\leq
T_{BG}\leq 91.3$ K. Above $91.3$ K, the behavior is indicative of a vortex
liquid. A linear response is observed for $h_{a}\leq 25$ mOe, but the
dissipation grows rapidly approaching an ohmic response ($\varepsilon =1$)
at $T\sim 91.5$ K (see Figure \ref{fig4}). However, at high $h_{a}$ we still
observe a nonlinear response, \textit{i.e.}, we cannot access experimentally the
unpinned (ohmic) liquid region. This is probably because at our low $H_{dc}$
the unpinned liquid regime only occurs in a very narrow temperature range
below $T_{c}$.

Let's now analyze the glassy phase of the dynamic diagram in terms of
existing models for vortex motion. According to the simplest scenario\cite
{Vander2}, the crossover between Campbell and Bean critical regimes should
occur at a field $h_{a}^{\ast }(T)$ where the range of field penetration in
the later (which is proportional to $h_{a}$) becomes larger than the range
of field penetration in the former (which is independent of $h_{a}$). For a
longitudinal geometry (e.g., a slab or cylinder in a parallel field) this
leads to the simple crossover condition $\Lambda _{c}(h_{a},T)\sim \lambda
_{R}(T)$ but, in our case, a meaningful comparison should involve the actual
field penetration ranges rather than the penetration depths. The practical
criterion that we have adopted is to consider that the range of field
penetration in the linear and critical regimes is the same when $\chi
^{,}=-1/2$ in each case. This crossover should occur when $\left( \sqrt{%
(2R/\delta )}/3\right) \Lambda _{c}(T)\sim \lambda _{R}(T)$, \textit{i.e.}, according
to Eq. (\ref{lambdaBean}), when

\begin{equation}
\lambda _{R}(T)\sim \frac{c}{4\pi }\left[ \frac{1}{3}\sqrt{\frac{2R}{\delta }
}\right] \frac{h_{a}^{\ast }(T)}{J(T)}  \label{hac.crossover}
\end{equation}

We can use Eq. (\ref{hac.crossover}) to estimate the crossover field $%
h_{a}^{\ast }(T)$ from $\lambda _{R}(T)$ and $J(T)$. We emphasize that $%
\lambda _{R}(T)$ and $J(T)$ are obtained in totally independent ways; the
former from the linear regime data (section III.C.1) and the latter
from the high $h_{a}$ data using the geometrical procedure described in 
section III.C.3. The result is shown in Figure \ref{fig7}: the
theoretical expectation for the crossover field $h_{a}^{\ast }(T)$ coincides
very well with the line $h_{a}^{c}(T)$ that indicates the formation of a
Bean critical state. This good agreement supports the validity of the basic
concept, \textit{the critical state develops when the Bean penetration range
becomes longer than the Campbell penetration range.}

We now discuss the origin and nature of the large intermediate regime. The
observation of a nonlinear response at current densities well below $%
J_{90KHz}$, the current density flowing in the Bean critical state,
indicates that these nonlinearities are associated with intra-valley
oscillations. This can be understood by considering the shape of the pinning
potential, that at these high temperatures is rather shallow. For the
present crystal we found in section III.C.2 that nonlinearities
develop when the amplitude of the oscillations at the sample border reaches
a value $u(\rho =R)=r_{l}\sim 50$ {\AA}. We associate $r_{l}$ with the range
where the pinning potential $\varepsilon _{p}\left( r\right) $ can be well
approximated by a parabola, which is smaller than the total pinning range.
This picture explains the smooth nature of the crossover from linear to
nonlinear response at $h_{a}^{l}(T)$.

As was discussed in section I.C., in the temperature and field range
of our ac measurements we are deep into the collective pinning regime. This
is in agreement with the conclusion of section III.C.3. However, the
linear response in the Campbell regime is well described by a field
independent Labusch parameter characteristic of single vortex pinning 
(section III.C.1). This apparent contradiction is easily resolved when the
size of the vortex displacements involved in each case are taken into
account. The linear response corresponds to oscillations that are uniform
along the field direction, whose amplitude $\left( \leq 50{\AA} \right) $ is
much smaller than the vortex lattice parameter $a\sim 2100$ {\AA}. In this
range of displacements the energy variation sensed by a vortex due to the
elastic interaction with the neighbor vortices is negligible as compared to
the variation of the pinning energy, and thus the response is solely
determined by the interaction of a single vortex with the tracks. We note,
however, that the transverse localization length is much longer than the
defect separation, thus each vortex is collectively pinned by many defects.
The amplitude of the oscillations refers to the vortex center of mass, and
the effective pinning potential arises from the contribution of more than
one track. In the critical state regime at high amplitudes, on the other
hand, the inter-valley vortex excitations involve transverse displacements
of the order of the defect separation or larger, and consequently the
vortex-vortex interactions play a fundamental role in the response.

\bigskip


\subsection{Dynamic regimes in tilted vortices.}

In this section we use the methods described above for aligned CD, to
complete the study of the angular dependence presented in section III.B with
a detailed analysis of the AC dynamic regimes at various orientations of $\bf{H_{dc}}$.
We will use the same irradiated crystal (sample A, $\Theta
_{D}=30^{\circ}$). The angles selected are representative of the 
various angular regions: $\bf{H_{dc}}$ forming an angle very small relative to
defects ($\Theta \sim 34^{\circ}$, $\varphi \sim 4^{\circ}$ ), $\bf{H_{dc}}$ along the
symmetric direction of tracks relative to the $c$ axis ($\Theta \sim -30^{\circ}$, 
$\varphi \sim -60^{\circ}$), and $\bf{H_{dc}}$ in the angular region where the
screening fall down ($\Theta \sim -70^{\circ}$ , $\varphi \sim 80^{\circ}$ ). For
comparison, we also include the response with aligned vortices ($\Theta \sim
30^{\circ}$, $\varphi \sim 0^{\circ}$ ) and the analogue response of the virgin
crystal (sample B) for $\Theta \sim 30^{\circ}$. In all cases, the response at
very low AC fields is linear, and it becomes non-linear above a threshold AC
field $h_{a}^{l}(\Theta ,T)$. In the following subsection we compare the
results obtained in the linear regime for each one of the chosen angles, and
later we analyze the non-linear response.

\subsubsection{Linear regime}

We have used the procedure explained in section III.C.1 to obtain
the linear real penetration depth and the Labusch constant in the linear
regime. To analyze the $B$ dependence of $\alpha _{L}$, we performed measurements for
several values of DC fields along the chosen directions in the linear regime,
and we calculated the corresponding $\lambda _{R}^{2}(\chi ^{,})$. We have 
observed qualitative differences in the behavior in the various angular ranges.

%
\begin{figure}[htb]
\centering
\includegraphics[angle=0,width=100mm]{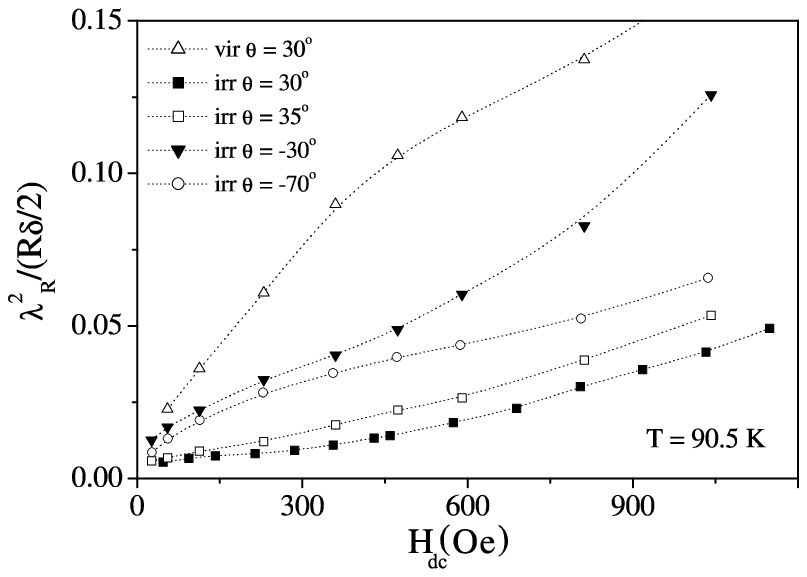}
\caption{{\protect\small Field dependence of the square of the dimensionless
linear real penetration depth in an irradiated sample at a fixed temperature
for different angles $\protect\Theta $. A curve of a virgin sample is also
shown.}}
\label{fig8}
\end{figure}

In figure \ref{fig8} various curves of $\lambda _{R}^{2}(B)$ for the chosen
values of $\Theta $ are compared at $T=90.5$ K. We can see that the
qualitative $\lambda _{R}^{2}(B)$ behavior observed when $\bf B$ is aligned 
with the tracks still holds if the field is tilted by a few degrees \
($\Theta =34^{\circ}$), but it changes notably for larger tilts, and it
is also very different in the non-irradiated sample. In the last two cases, no
linear dependence in $\lambda _{R}^{2}$ vs. $B$ at low fields is observed,
indicating that a Labusch constant independent of field does not exist. It
can be seen that the Campbell penetration length $\lambda _{C}$ is much
shorter when $\bf{H_{dc}} \parallel$ CD than for angles far away (4
or 5 times) and than for the non-irradiated sample (10 times). For fields
higher than $B_{\phi }/2$, the ratio between the Campbell lengths in
different orientations decreases.

Furthermore, we observe that the response at $\Theta =-30^{\circ}$ in the
irradiated sample is different to the one observed at $\Theta =30^{\circ}$ in
all the range of measured fields; this means that up to $H_{dc}\approx
2B_{\phi }$ the presence of defects is still important. A noteworthy result
is that, for low DC fields, the responses at $-70^{\circ}$ and $-30^{\circ}$ are
very similar. For $\Theta =-30^{\circ}$ near $H_{dc}=B_{\phi }/2$, the curve
starts to increase more swiftly, and both responses can be clearly
distinguished. This fact indicates that even at an angle $\varphi $ $\sim
-60^{\circ}$ from the tracks direction, the occupation of defects play an
important role.

The above experimental results can be qualitatively understood as follows.
In section III.B. we concluded that vortices are partially
accommodated in the defects for all the chosen angles. The restitutive
constant $\alpha _{L}$ is higher in the vortex segments that are
accommodated in the columnar defects. For this reason, while the Campbell
regime remains individual, the average Labusch constant will increase with
the length of pinned segments, \textit{i.e.} when the angle $\varphi $ (relative to
defects) decreases. For large $\varphi $ the situation is more complex: the
fraction pinned by the tracks is very small (it can be seen that $\alpha
_{L} $ for the non-irradiated sample at $30^{\circ}$ is just half the value
observed at $-30^{\circ}$ in the irradiated sample). Moreover, the pinning force
for small displacements is comparable to that induced by neighbor vortices
and $\alpha _{L}$ is no longer independent of $B$.

\bigskip

\subsubsection{Nonlinear response}

\begin{figure}[htb]
\centering
\includegraphics[angle=0,width=150mm]{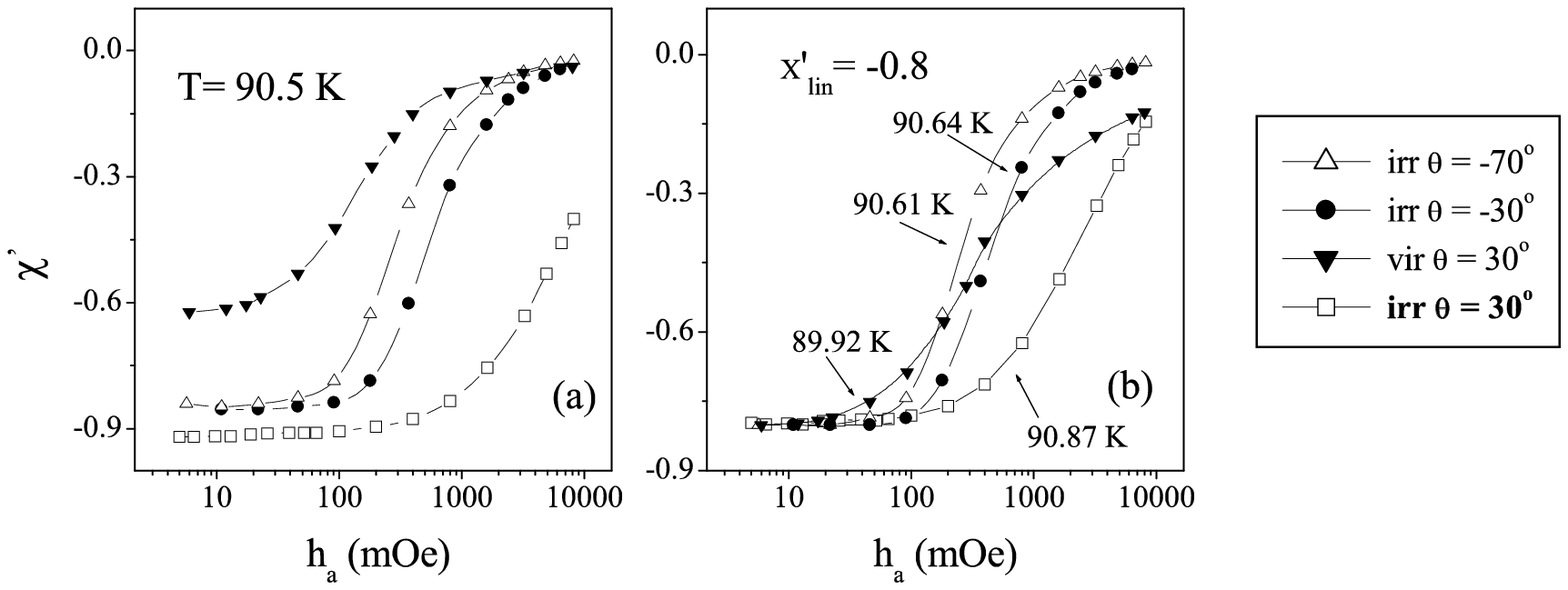}
\caption{{\protect\small Linear and nonlinear behavior at different angles $%
\protect\Theta $ as a function of the ac field. (a) Comparison of $\protect%
\chi ^{,}(h_{a})$ curves at the same temperature. (b) Comparison at the same
linear real penetration depth. Symbols correspond to both panels.}}
\label{fig9}
\end{figure}

Measurements in all the available range of $h_{a}$ have been performed. The
first notable observation is that, in the non linear regime, the angular
dependence of the AC response is much more pronounced than in the linear regime. 
For large angles relative to the CD, as soon as the linear behavior
is lost the $\chi (T)$ curves get wider in temperature and the dissipation
peaks notably increase. These facts are resumed in figure \ref{fig9}, where
the non linear behavior at different angles and their comparison with the
non-irradiated sample are shown. In figure \ref{fig9}(a), various $\chi
^{,}(h_{a})$ curves at the same temperature ($90.5$ K) for different angles
are plotted, while in figure \ref{fig9}(b) all the curves included have the
same $\lambda _{R}$ (and therefore they were obtained at different
temperatures).

%

It can be observed that the AC field $h_{a}^{l}$ at which loss of linearity
occurs is lower when the pinning due to the CD is less efficient: in the
non-irradiated sample (solid down triangles in the figures), $h_{a}^{l}$ is
almost one order of magnitude lower than in the case of the irradiated
sample with field aligned to defects (open squares in the figures). When 
$\bf{H_{dc}} \parallel$ CD, departure from linearity is smooth, while for
large $\varphi $ angles (solid circles and open up triangles in the figures)
the departure from linearity is much more abrupt.

\begin{figure}[htb]
\centering
\includegraphics[angle=0,width=80mm]{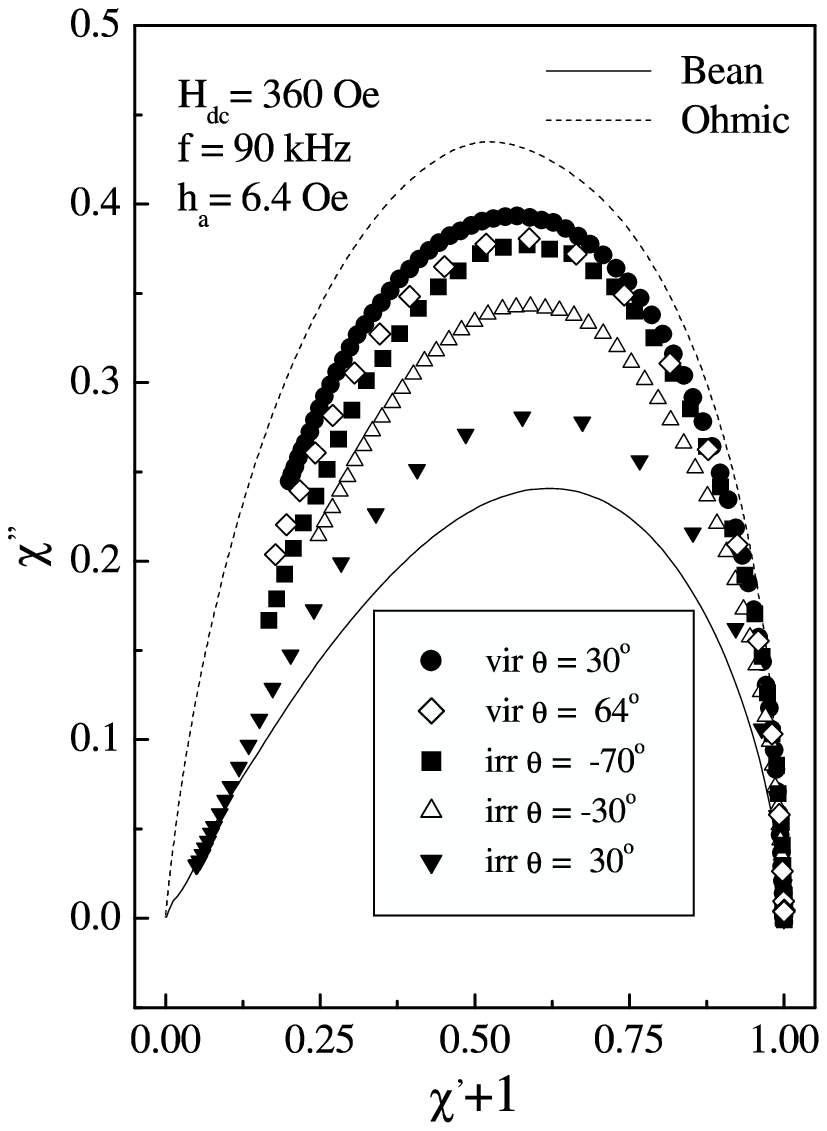}
\caption{{\protect\small Experimental curves $\protect\chi ^{,,}(\protect\chi
^{,})$ at a large $h_{a}=6.4$ Oe in the irradiated an virgin sample for
various angles. Lines are the theoretical calculation for a disk in an Ohmic
regime (dashed line) and in a Bean critical state (solid line).}}
\label{fig10}
\end{figure}

The dependence of $\chi $ on $h_{a}$ for the highest $h_{a}$ is also very
different. Whereas in the case of aligned columnar defects this function can
be explained in the framework of a critical state model, for large $\varphi $
angles this dependence decreases tending to a new linear regime. This
behavior is still more notable in the non-irradiated sample. Consistently
with the last observation the maximum of $\chi ^{,,}(T)$ approaches the
expected values\cite{brandtA} for the ohmic regime ($\chi _{\max }^{,,}\sim 0.44$ for a
disk in transversal geometry). The last remark is displayed in figure \ref
{fig10}. There we compare experimental points of $\chi ^{,,}$ vs. $\chi
^{,}+1$ obtained at different angles at $h_{a}\sim 6.4$ Oe, with those
calculated for a disk both in the Bean critical state (continuous line in
the figure) and in the ohmic regime (dashed line). While in the irradiated
sample the response at $\Theta =30^{\circ}$ is similar to the expectation for a
critical state, for angles far away from the defects it tends to the ohmic
behavior. Notice that, as well as $\chi _{\max }^{,,}$ increases, $\chi
^{,}(\chi _{\max }^{,,})$ also tends to the expected value in a linear ohmic
regime.

%

As another test to close this picture, we tried to check whether a critical
state regime is established or not, applying the procedure presented in
section III.C.3 to data obtained at $\Theta =-70^{\circ}$. A satisfactory
solution was not attained, proving that those data are not consistent with
the critical state model.

A possible explanation for this behavior is the following: When $\varphi$ is
large, the effective critical current density $J_{\omega }$ is much lower
than in the case in which $\bf {H_{dc}}$ is aligned with the tracks. In fact, it
is so much reduced that a critical state cannot be established: as soon as
the linearity is lost, and the vortex displacements become mainly determined
by creep mechanisms across the pinning centers, the AC field penetration
length increases very much, and becomes comparable to the flux flow skin
depth. In this condition, a nonlinear regime in which pinning forces,
activated mechanisms and viscous losses contribute significantly to the
vortex motion is established.

Finally, another interesting consideration arises from figure \ref{fig9}, as
it allows us to re-analyze the angular dependence of $\chi (\Theta )$
previously shown in section III.B, figure \ref{fig2}. The angle
beyond which there is an abrupt increase in the AC field penetration in the
left side of figure \ref{fig2} ($\Theta \approx -60^{\circ}$), corresponds to
the angle for which, at the particular AC field value $h_{a}=0.2$ Oe, the
system changes from a nearly linear regime to a non linear one. This fact
can be easily verified crossing the curves $\chi ^{,}(h_{a})$ at $h_{a}=200$
mOe in figure \ref{fig10}(a). By performing the same procedure at lower $%
h_{a}$ we can observe that, if angular measurements were made at lower AC
fields, a smooth angular dependence would extend to orientations closer to
the $ab$ planes. On the other hand, if we perform the procedure for $%
h_{a}=400$ mOe, the abrupt increase of AC field penetration will appear
before $\Theta =-30^{\circ}$. For even higher values of $h_{a}$, the region
drastically affected by the CD will only involve the peak close to the
defects, as it has been reported previously \cite{herbsommer98}. From all
this considerations, we point out that the apparent angular region notably
affected by the CD is $h_{a}$ dependent. Thus, results of angular
measurements performed using only using AC amplitude, without characterizing
the dynamic regimes involved, should be interpreted with caution.

\section{Conclusion}

High temperature superconductors with correlated disorder provide a fascinating framework for the exploration of novel ideas on vortex physics. In addition, our understanding of this topic will be essential for the optimization of YBCO superconducting wires for technological applications, as vortex motion in those materials is also dominated by correlated pinning, arising in that case from fabrication-dependent microstructures.

We have explored in detail the angular dependent vortex dynamics in type II superconductors with aligned columnar defects introduced by irradiation with very energetic heavy-ions. We have used dc magnetization measurements deep in the vortex solid phase, and ac susceptibility near the solid-liquid transition. We have shown that aligned columnar defects are an excellent tool to test models for vortex dynamics, particularly if they are tilted with respect to the crystallographic axes, so their effects can be easily distinguished from those arising from mass anisotropy, sample geometry, twin boundaries and intrinsic pinning. This allows us, for instance, to use the uniaxial pinning of the columnar defects as a probe to determine the orientation of the vortices inside a bulk material, which in general is different from the orientation of the applied fields. 

In some aspects we have found an excellent agreement with the theoretical expectations of the Bose-glass model. The field dependence of the lock-in angle follows remarkably well the $1/H$ prediction over the whole temperature range of our measurements. In turn, the temperature dependence of the lock-in angle gives strong support to the concept of an effective pinning energy dominated by the entropic smearing effect. On the other hand, both our ac and dc results show that columnar defects produce effective pinning over a wide angular range, and that correlated pinning dominates the scenario for all field orientations. One consequence of this is the existence of a rich variety of vortex staircases, with segments locked into different correlated structures. 

The complexity of the picture is even larger when the high temperature dynamics is taken into account. When the magnetic field is aligned with the tracks the ac susceptibility at low ac excitations exhibits a linear response arising from vortices oscillating inside the tracks, which is characterized by a Labusch parameter independent of the dc field. This indicates that the response is dominated by the individual vortex-track interaction, even though the critical current in the same temperature and field range is determined by collective pinning mechanisms. The solution of this paradox is related to the fact that the characteristic size of the vortex displacements in each case is very different. Finally, the ac response at different angles indicates that the characteristics of the dynamic regimes and their extension in the temperature vs ac field plane are strongly dependent on the orientation of the vortices with respect to both the columnar defects and the crystalline structure. 

\section{Acknowledgements}

We are pleased to thank the Atomic Energy Commission of Argentina, where all the measurements presented in this review have been performed. We also want to thank the CONICET of Argentina for financial support. The experimental results described in this review were obtained in collaboration with H. Lanza, G. Nieva, P. Levy, M. Avila, D. Niebieskikwiat, S. Candia and D.
Casa. We acknowledge many valuable discussions with V. Bekeris, G. Blatter, F. de la Cruz, D. Lopez, J. Guimpel, A. Herbsommer, L. Krusin-Elbaum, J.R. Thompson, and S. Valenzuela.

\end{document}